# Identifying Concerns When Specifying Machine Learning-Enabled Systems: A Perspective-Based Approach

Hugo Villamizar[1*], Marcos Kalinowski[1],
Hélio Lopes[1], Daniel Mendez[2]

[1*]Pontifical Catholic University of Rio de Janeiro, Rio de Janeiro, Brazil.
[2]Blekinge Institute of Technology, Karlskrona, Sweden.

*Corresponding author(s). E-mail(s): hvillamizar@inf.puc-rio.br;
Contributing authors: kalinowski@inf.puc-rio.br; lopes@inf.puc-rio.br;
daniel.mendez@bth.se;

**Abstract**

Engineering successful machine learning (ML)-enabled systems poses various challenges from both a theoretical and a practical side. Among those challenges are how to effectively address unrealistic expectations of ML capabilities from customers, managers and even other team members, and how to connect business value to engineering and data science activities composed by interdisciplinary teams. In this paper, we present *PerSpecML*, a perspective-based approach for specifying ML-enabled systems that helps practitioners identify which attributes, including ML and non-ML components, are important to contribute to the overall system's quality. The approach involves analyzing 59 concerns related to typical tasks that practitioners face in ML projects, grouping them into five perspectives: system objectives, user experience, infrastructure, model, and data. Together, these perspectives serve to mediate the communication between business owners, domain experts, designers, software and ML engineers, and data scientists. The creation of *PerSpecML* involved a series of validations conducted in different contexts: (i) in academia, (ii) with industry representatives, and (iii) in two real industrial case studies. As a result of the diverse validations and continuous improvements, *PerSpecML* stands as a promising approach, poised to positively impact the specification of ML-enabled systems, particularly helping to reveal key components that would have been otherwise missed without using *PerSpecML*.

# 1 Introduction

Contemporary advances in Machine Learning (ML) and the availability of vast amounts of data have both given rise to the feasibility and practical relevance of incorporating ML components into software-intensive systems. In this paper, we refer to them as ML-enabled systems. These systems have their behavior dictated by data instead of relying on explicitly defined rules. In other words, data replaces code to some extent. This shift from engineering purely conventional software systems to ones which have ML-components woven-in poses new challenges from the viewpoint of software engineering (SE); for instance, challenges related to covering quality properties such as fairness and explainability [21], or challenges related to collaboration and mismatched assumptions in ML projects given the required multidisciplinary teams [31, 37]. These particularities typically demand extra effort to successfully develop ML-enabled systems. It is, therefore, not surprising to us that Gartner reports only 53% of ML projects to make it into production [19].

Within SE, Requirements Engineering (RE) is, in simple terms, the discipline that is meant to effectively translate stakeholder needs into requirements, constraints, and other information that defines what software systems should do under which conditions [13]. Due to the communication and collaboration-intensive nature, as well as inherent interaction with most other development processes, RE can provide the very foundation to address several of the challenges of building ML-enabled systems [27]. For example, when developing ML models, we need to identify relevant and representative data, validate models, and balance model-related user expectations (*e.g.*, accuracy versus inference time); just as in RE for conventional software systems where we need to identify representative stakeholders, validate specifications with customers, and address conflicting requirements.

This has also caught a new level of interest by the research community trying to better understand how RE techniques can be extended and what challenges need to be solved to reliably build ML-enabled systems [12]. Literature has shown that identifying quality metrics beyond accuracy, their specification, and understanding how they can be analyzed are not well-established yet in ML contexts [1, 2, 40, 45]. In fact, a recent roadmap for the future of SE [8] emphasizes that existing RE methods will need to be expanded to decouple ML problem and model specification from the system specification. On a more practical side, outside of BigTech companies with lots of experience, there is a focus on training more accurate ML models and their deployment, but rarely on the entire system including ML and non-ML components (*e.g.*, how data is collected, how mistakes are dealt). This may lead to incomplete specifications of ML-enabled systems [23, 30], leaving most decisions to be made by data scientists [31, 48].

In order to help addressing these issues, we present *PerSpecML*, an approach for specifying ML-enabled systems that involves analyzing 59 concerns grouped into five perspectives: system objectives, user experience, infrastructure, model, and data. Together, these perspectives serve to mediate the communication between business owners, domain experts, designers, software and ML engineers, and data scientists. We created *PerSpecML* by following a technology transfer model [20], which is recommended to foster successful transfer of technology from research to practice [50].

Throughout this process, we participated in real ML projects of a research and development (R&D) initiative [26], conducted a literature review on RE for ML [45], created a catalogue with an initial set of concerns [46], and proposed a candidate solution for specifying ML-enabled systems [47]. In this paper, we iteratively evaluate and improve [46, 47] by conducting three studies in different contexts: (i) in an academic validation involving two courses on SE for data science, (ii) with practitioners working with ML-enabled systems in an R&D initiative, and (iii) in two real industrial case studies conducted with a Brazilian large e-commerce company.

The iterative validations and continuous improvements result in *PerSpecML*, our approach for specifying ML-enabled systems, and collectively corroborated its potential as a comprehensive tool for guiding practitioners in collaboratively designingML-enabled systems, enhancing their clarity, exploring trade-offs between conflicting requirements, uncovering overlooked requirements, and improving decision-making. Furthermore, we found that the participants involved in the validations gradually improved their perception of *PerSpecML*'s ease of use, usefulness, and intended to use.

The remainder of this paper is organized as follows. Section 2 presents the background and related work. In Section 3, we detail how we conceive, evaluate and evolve *PerSpecML*. In Section 4, we present *PerSpecML* and details its elements. In Sections 5 and 6, we describe the evaluation in academia and with industry representatives. Section 7 reports on industrial case studies. Section 8 and Section 9 raises potential threats to validity and discusses our research findings. By last, in Section 10, we conclude the paper.

## 2 Background and Related Work

This section introduces a background on the core essence of ML and presents particularities and challenges when engineering ML-enabled systems that RE may address. We also describe related work.

### 2.1 ML in a Nutshell

ML is the study of computer algorithms that explores data to determine the best way to combine the information contained in the representation (training data) into a model that generalizes to data it has not already seen [35]. These systems, unlike non-ML, base its behavior on external data instead of explicitly programming hard rules. However, data may not be adequate and lead to bad outcomes. The output of the ML model is a prediction, sometimes surprisingly accurate and sometimes surprisingly inaccurate. When an ML model is integrated into a functional system, it becomes an ML-enabled system. This supposes a change in the way of designing, developing and testing these type of systems.

### 2.2 Quality of ML-Enabled Systems

Assuring the quality of ML-enabled systems is essential since these systems are increasingly becoming part of our daily life. However, this is not an easy task. Their quality goes beyond ML model performance metrics such as accuracy, precision or recall.

Typically, these ML model performance metrics comprises the primary goal of data scientists during ML model development. A good ML-enabled system is one in which the learning improves over time, particularly when the learning improves by getting feedback from users. This implies taking care of not only data and models, but also business context, user experience, infrastructure and integration of several services. When designing an ML-enabled system is important to understand the constraints on its operation. For example, where will the model run? What data will it have access to? How fast does it need to be? What is the business impact of a false positive? A false negative? How should the model be tuned to maximize business results? An ML model is just one component of an ML-enabled system as a whole. There is an incredible amount of work to be done between the development of an ML model, the incorporation of it into a system and the eventual sustainable customer impact [6, 23, 30]. Thinking about possible strategies to address these concerns increases the chance of designing and development an ML-enabled system that meets customer's needs, and can avoid often costly problems later.

## 2.3 RE for ML-Enabled Systems

Requirements Engineering (RE) constitutes approaches to understand the problem space and specify requirements that all stakeholders agree upon. As such, it is concentrates on understanding what the actual problem is, what needs towards a system result and how to resolve potential conflicts, and it is thus characterized by the involvement of interdisciplinary stakeholders and often resulting in uncertainty [49]. RE is often considered a crucial and challenging stage of any software project. Indeed, most of the problems in software systems with and without ML components come from poor requirements rather than faulty implementation. In this line, Kästner [27] states that an ML model can be seen as a specification based on training data since data is a learned description of how the ML model shall behave. This means that the learned behavior of an ML-enabled system might be incorrect, even if the learning algorithm is implemented correctly.

Practitioners argue that the incorporation of ML implies addressing additional qualities, setting more ambitious goals, dealing with a high degree of iterative experimentation, and facing more unrealistic assumptions [36]. It is therefore reasonable to assume that handling and resolving validation problems is (or should be) in scope of the role of a requirements engineer. We further argue that investing in RE can help to identify and mitigate problems early on. Nevertheless, establishing RE may be difficult due to the lack of guidance, tools, and techniques to support the engineering of ML-enabled systems [1, 45]. It is not surprising that ML-enabled systems are rarely built based on comprehensive specifications [31, 32] and that RE is seen by practitioners as the most difficult phase in ML projects [23].

In the last years, the literature on RE for ML has focused on issues with data requirements [9], process of data-driven projects [48], challenges of addressing non-functional requirements and particularities of certain quality attributes such as explainability, transparency and fairness [11, 21, 34]. Despite the important contributions in the field so far, the importance of specifying ML components in a way that customers can understand and analyze it to make adequate decisions is too often overlooked [15], and only a limited number of studies have looked into how to specify and document requirements for ML-enabled systems [1, 2, 40, 45]. For instance, Berry [7] states that the measures used to evaluate a learned machine, the criteria for acceptable values of these measures, and the information about the ML context that inform the criteria and trade-offs in these measures, collectively constitute the requirements specification of ML-enabled systems.

## 2.4 Related Work

We subsequently highlight research that has investigated what quality attributes should be analyzed and how practitioners can specify and document requirements for ML-enabled systems. We further take a more holistic RE perspective where an ML model is merely part of a larger ML-enabled system.

Dorard [16] proposed a management template for ML, also known as ML canvas, that can be used to describe how ML systems will turn predictions into value for end-users, considering elements such as problem definition, data collection and preparation, feature engineering, model selection, evaluation metrics, deployment, and monitoring. This is probably the most spread approach for documenting ML-enabled systems given its simplified representation. However, this can be seen as a limitation since ML canvas may not capture all the intricate details and complexities of real-world projects, leading to potential oversights or gaps in the analysis. We seek to bridge these gaps with *PerSpecML* by focusing on five different perspectives covering technical aspects and broader contextual concerns such as ethical considerations, legal constraints, and business implications, which can be crucial in real-world implementations.

Rahimi *et al.* [41] discussed on ideas for extracting and visualizing safety-critical requirements specifications and how a self-driving car would recognize pedestrians. The authors describe how RE can be useful to better understand the domain and context of a problem and how this helps to better select a high-quality dataset for model training and evaluation. We are aware that identifying gaps in the associated dataset and the constructed ML model is essential to improve the overall quality, fairness, and long-term effectiveness of the ML-enabled system, but at the same time other external components such as those related to the operation (*e.g.*, data streaming) play an important role and can make the difference between an ML-enabled system that fits customer's needs and one that doesn't.

In an effort to model a representation of data-driven systems, several works have been proposed. For instance, Chuprina *et al.* [10] presented an artefact-based RE approach that encompasses four layers: context, requirements, system, and data. While the context specification captures the operational environment of a system, the requirements specification covers the user-visible black-box behaviour and characteristics such as explainability, transparency and ethics. On the other hand, the system specification defines the solution space and considers the system in a glass box view. The data-centric layer captures artifacts such as training and test datasets, and verifying algorithms. Similarly, Nakamichi *et al.* [38] proposed a requirements-driven model to determine the quality attributes of ML-enabled systems that covers perspectives such as environment/user, system/infrastructure, model, data and quality characteristics.

Despite the important contributions of these works, we found some limitations when compared to *PerSpecML*. Firstly, our intention is to be more specific, including more fine-grained attributes for each layer/perspective and modeling their relationships so that practitioners can have a complete view of the ML context and the software system as a whole. Secondly, we detail ML-related concerns that we faced in practice that were not considered as part of their proposals, such as concerns related to business requirements and user experience, which in our context showed being important for the success of ML-enabled systems.

Another study we consider relevant is one conducted by Nalchigar [39]. They reported on an empirical study that evaluates a conceptual modeling framework for ML solution development for the healthcare sector. It consists of three views consumed by business people, data scientists, and data engineers. The business view shows how business goals are refined into decision goals and question goals, and how such questions can be answered by ML. The analytic design view models a solution in terms of algorithms, non-functional requirements and performance indicators. Lastly, the data preparation view conceptualizes the design of data preparation tasks in terms of data tables, operations, and flows. We also find this work as relevant as the previous ones, but we believe that other views related to the operation of ML-enabled systems such as infrastructure and user experience must be considered to support the activities of practitioners such as software and ML engineers, and designers.

More recently, Siebert *et al.* [43] presented a formal modelling definition for quality requirements in ML-enabled systems that allows to identify attributes and quality measures related to components such as model, data, system, infrastructure and environment. We consider this work strongly related to ours. For instance, the authors discusses quality attributes of an ML-enabled system beyond the ML components, just as *PerSpecML* proposes. It is also explicit about considering multiple perspectives: of the entire system, and of the environment the system is embedded it. As a key difference between the works, we provide a diagram that summarizes the perspectives, the quality attributes/concerns, and shows their relationships. This seeks to facilitate effective communication and collaboration among stakeholders, provide a visual representation that can be easily understood by technical and non-technical team members, capture and document various aspects of the ML-enabled system's design, and support analysis and verification activities.

Similarly, Maffey *et al.* [33] proposed MLTE, an initial framework to evaluate ML models and systems that provides domain-specific language that teams, including model developers, software engineers, system owners, can use to express model requirements, an infrastructure to define, generate, and collect ML evaluation metrics, and the means to communicate results. While MLTE defines a general measurable process to evaluate ML systems, our proposal differs by going a step back and pointing out typical concerns involved when setting objectives and defining key components of ML-enabled systems. We see MLTE and *PerSpecML* as tools that can complement each other by supporting practitioners from different angles, since they share the same purpose of early addressing practical problems faced by multidisciplinary teams throughout the ML development process.

# *3* Methodology for Conceiving *PerSpecML*

In this section, we describe the process we followed to design and evaluate *PerSpecML* based on the technology transfer model introduced by Gorschek *et al.* [20]. We used this model since our research method involved evaluations in both academia and industry with the aim of scaling the proposal up to practice, for which this model is recommended [50]. This mix of evaluations provides an opportunity to gather user feedback and incorporate it into the solution design. By involving stakeholders and practitioners in the evaluation process, we gathered valuable insights about their experience, needs, and preferences. This feedback informed iterations and refinements of the solution, making it more user-centric and aligned with actual user requirements. Fig. 1 outlines the seven steps of the model, which we will describe sequentially hereafter (while following the terminology of the transfer model).

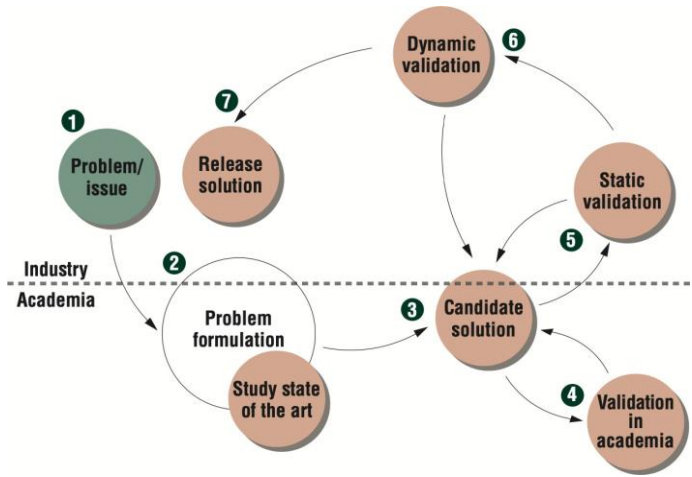

**Fig. 1** Technology transfer model proposed by Gorschek *et al.* [20]

## 3.1 Step 1: Identify Improvement Areas Based on Industry Needs

We followed the principle of constructivism [18] that advocates that a person needs to understand how something works before exploring the different ways to construct solution proposals. During the last four years, the first author has participated in research and development (R&D) projects designing and developing ML-enabled systems. These projects involve different types of ML tasks (*e.g.*, supervised and unsupervised learning, computer vision) and algorithms (*e.g.*, decision trees, logistic regression, neural networks).This experience allowed us to assess current practices, observing domain and business settings, understand typical industry needs for ML-enabled systems, and issues related to their development. More specifically, we identified i) how important the domain and business settings are to align the stakeholder needs, requirements, and constraints with the engineering and data science activities ii) interdisciplinary teams typically involved in ML projects, and iii) the lack of tools and documents that can capture key components when specifying ML-enabled systems.

## 3.2 Step 2: Formulate a research agenda

In order to better define the problem and gain more insights into existing solutions and what needs to be created, we conducted a systematic mapping study on RE for ML [45], analyzed later literature reviews [1, 2, 40] and took advice from an industry-oriented publication based on more than a decade of experience in engineering ML-enabled systems [22]. Here, we identified, for instance, i) additional quality attributes of ML-enabled systems that practitioners should analyze ii) the lack of studies focused on identifying key components of ML-enabled systems that may later be specified, and iii) the lack of studies evaluated in practice to validate its effectiveness, feasibility and gather user feedback.

## 3.3 Step 3: Formulate a Candidate Solution

After observing and gathering experience from real-world ML projects and reviewing the literature, we decided to focus on the creation of a candidate solution that can support the design of ML-enabled systems. As a first step, we proposed a catalog of 45 concerns to be analyzed by practitioners with the aim at identifying key components of ML-enabled systems [46]. The initial set of concerns were evaluated in a focus group with practitioners with different levels of experience of a R&D initiative, more specifically, three data scientists, two developers and three project leads. Their feedback was positive as they perceived the catalog of concern as prominent, and allowed us to identify initial improvements. Fig. 2 shows the catalog.

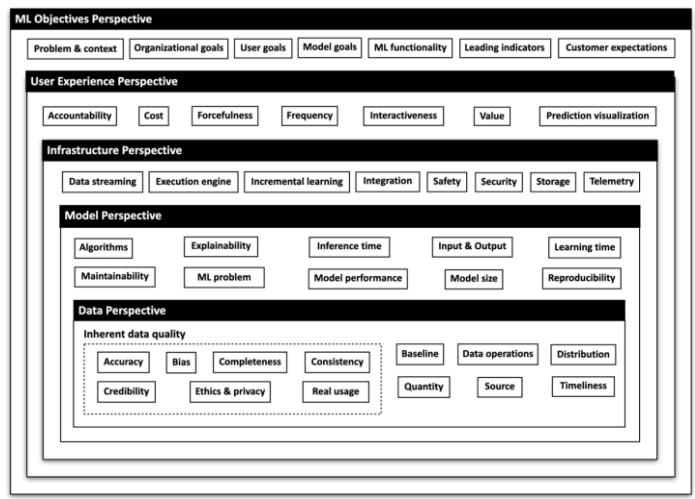

**Fig. 2** Initial catalog of concerns [46]

Therefrom, we used this catalog to create a candidate solution for specifying ML-enabled systems [47]. This candidate solution modeled the concerns in a structured manner by proposing a diagram that categorizes the concerns into perspectives, pointing out relationships and stakeholders involved in the analysis of the concerns. The purpose was to capture essential information about the desired functionality, components, and constraints of the ML-enabled system. Fig. 3 shows the diagram we proposed in a first effort to specify ML-enabled systems.

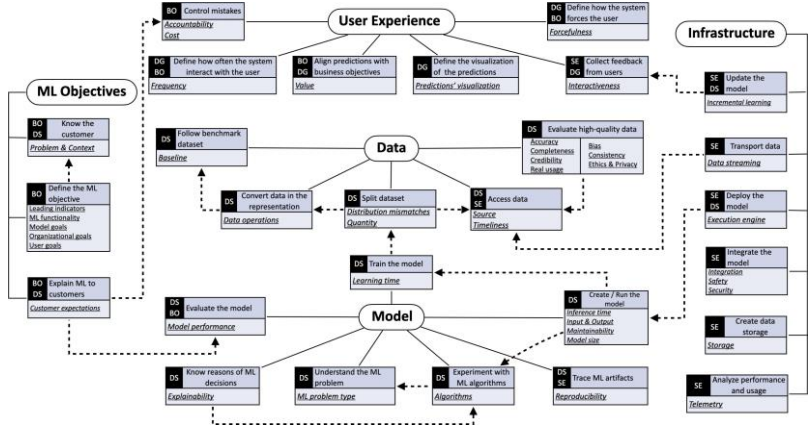

**Fig. 3** Initial diagram for specifying ML-enabled systems [46]

In this paper, we iteratively improve this candidate solution by conducting three different evaluations that are briefly described hereafter. The resulting approach, which we baptized *PerSpecML*, is detailed in Section 4.

## 3.4 Steps 4, 5, and 6: Evolution and Transfer Preparation through Validation

The goal of these steps was to refine the candidate solution towards its industry-readiness. In order to accomplish this goal, we conducted three evaluations in different contexts, as suggested by the technology transfer model [20]: (i) with students from two courses on SE for data science specifying an ML-enabled system for a toy scenario (validation in academia), (ii) with practitioners working in a R&D initiative discussing specifications of ML-enabled systems built retroactively with stakeholders of real projects (static validation), and (iii) in two industrial case studies conducted with an e-commerce company, specifying real ML-enabled systems from scratch using the approach (dynamic validation). Note that, according to [20], the terminology 'static' refers to evaluating the candidate solution off-line, involving industry participants and real artifacts, but not as part of a real project life-cycle activity, which is the 'dynamic' one. With these iterative validations we seek to ensure early issue detection, user satisfaction, continuous improvement, adaptability and overall confidence in the final solution. Details on the validations are provided in Section 5, 6, and 7.

## 3.5 Step 7: Release the Solution

*PerSpecML*, which is presented in the next section, is now being adopted within the R&D initiative involved in the static validation to specify their ML-enabled system projects. In addition, the approach has been successfully transferred to the data science team responsible for the two case study projects involved in the dynamic validation. At first, the team decided to limit *PerSpecML* to ML projects involving supervised learning tasks. The full adoption is pending results from other evaluations.

# 4 *PerSpecML*

In this section we present *PerSpecML*, a perspective-based approach for specifying ML-enabled systems that involves analyzing 59 concerns related to typical tasks that practitioners face in ML projects when defining and structuring these software systems. The concerns are grouped into five perspectives: system objectives, user experience, infrastructure, model, and data, providing a structured way to analyze and address different aspects of the ML-enabled system. Together, these perspectives align the activities between business owners, domain experts, designers, software and ML engineers, and data scientists. By using *PerSpecML*, practitioners are expected to be able to:

- **Enhance clarity:** Different stakeholders such as software engineers and data scientists may have varying goals, requirements, and concerns. Modeling perspectives and tasks helps to identify and explicitly represent these diverse viewpoints, ensuring a clear understanding of the ML-enabled system from multiple angles.
- **Foster collaboration:** Providing a perspective-based approach encourages collaboration and communication among stakeholders. It facilitates discussions and negotiations by providing a common structure to express and compare different viewpoints.
- **Identify trade-offs:** Perspectives and concerns enable the exploration of trade-offs between conflicting objectives and requirements. By explicitly modeling a high-level ML-enabled system workflow, practitioners can analyze the impact of design decisions on each perspective and make informed choices that balance different concerns.
- **Improve decision-making:** Understanding the tasks and concerns of both ML and no-ML components helps practitioners to evaluate and compare alternative solutions, enabling informed decision-making as the project progresses. ML projects are full of decisions that stakeholders must make.
- **Ensure completeness:** By considering multiple perspectives and concerns, practitioners can uncover hidden or overlooked requirements or risks. This helps in ensuring that the final ML-enabled system addresses the needs of all stakeholders and avoids potential pitfalls or shortcomings.

In the following, we detail each element of *PerSpecML* that we evolved throughout the iterative validations we conducted. We describe the stakeholders, the perspectives and their concerns, the relationship between them, and the two final artifacts that structure the above elements: the perspective-based ML task and concern diagram and the corresponding specification template. We also describe the logical flow for executing *PerSpecML*.

## 4.1 Stakeholders

Building successful ML-enabled systems requires a wide range of skills, typically by bringing together team members with different specialties [22, 28]. Taking a holistic system view is essential because ML expertise alone is not sufficient and even engineering skills to, for example, build pipelines and deploy ML models cover only small parts of the software system. We also need to be concerned about how to improve the experience of end-users in order to deal with unrealistic assumptions, and align business value to ML technical activities in order to cover business requirements. Given this, we seek *PerSpecML* to impact the work of business owners, domain experts, designers, software/ML engineers, data scientists and requirements engineers.

**Business owners (BO)** should understand what properties and components are essential to achieve the business objectives and be aware of the ML capabilities in order to set realistic goals and expectations. For instance, how to connect business objectives with ML outcomes? What is the real cost involved in maintaining an ML-enabled system? What team and skills are needed to successfully building ML-enabled systems?

**Domain experts (DE)** play an important role in accurately defining the problem in a way that aligns with real-world scenarios and requirements, ensuring that the ML-enabled system addresses the specific challenges and objectives of the domain. By collaborating closely with domain experts, other stakeholders can benefit from their in-depth knowledge and insights to define relevant features and data sources, and interpreting the results of the ML model in a meaningful context.

**Designers (DG)** collaborate to translate complex ML concepts and model outputs into intuitive and easy-to-understand interfaces that provide value to end users. For instance, where and how the ML outcomes will appear? how often it will appear? and how forcefully it will appear? A good user experience must be on the user's side and make them happy, engaged, and productive. Creating interactions with users to get feedback and grow learning is essential to ensure the quality of the ML model over time.

**Software/ML engineers (SE)** should understand how the entire system will interact with the ML model. They work on transforming the data scientists' research prototypes into ML-enabled systems that can handle large-scale data, ensure scalability, and meet performance concerns. For instance, what are the pros and cons of deploying an ML model as a back-end application or as a web service? online or batch predictions are enough to meet user demand?

**Data scientist (DS)** leverages their expertise in data analysis, statistical modeling, and ML algorithms to extract insights, develop ML models, and drive data-driven decision-making, but they should also understand the constraints these systems put on the ML models they produce. For instance, what quality properties the ML model should consider? What domain restrictions may apply? what should be the complexity of the ML model? and how should the ML model be tuned to maximize business results?

**Requirements engineers** collaborate closely with stakeholders to support the discussions between business owners, domain experts, and data scientists, and the development team, facilitating effective communication and understanding of project requirements. We seek to empower requirements engineers by using *PerSpecML* to identify and resolve conflicts often associated with ML projects. For instance, how much loss of accuracy is acceptable to cut the inference latency in half? can data scientists sacrifice some accuracy but offer better interpretability and explainability? One of the main benefits of applying RE for ML projects is to help balance these concerns.

## 4.2 Concerns

In SE, a concern typically refers to a specific aspect, interest, or issue that needs to be addressed or considered during the development and maintenance of a software system, consequently influencing its design, implementation and behavior. When designing ML-enabled systems and breaking them down into components, it is crucial to identify which attributes are important to contribute to the overall system's quality. Determining this requires a deep understanding of the system's goals, stakeholders' requirements, and the overall context in which the software will be used. In the case of ML components, the challenge is further amplified since it incorporates models that make predictions based on patterns and trends learned from data, which introduce unique considerations. All of these considerations, including ML components and deterministic (non-ML) components, become concerns for practitioners in charge of designing an ML-enabled system.

One of the main elements of *PerSpecML* are its concerns. In total, we identified 59 concerns including, for example, data streaming, model serving and telemetry when thinking on the operation of the ML-enabled system, and inference time, explainability and reproducibility when thinking on the development of the ML model. The concerns, that can be seen as quality attributes, came from i) own experiences of the authors of this work who have been actively participated in real ML projects, from ii) literature reviews on RE for ML that have researched both academia and industry, and from iii) practitioners who iteratively evaluated the concerns and recommended new ones to be considered. In *PerSpecML*, the concerns are part of tasks that stakeholders typically face throughout the development of ML-enabled systems.

## 4.3 Related Tasks Modeling

In *PerSpecML* we also focus on capturing and representing the tasks that should be performed by stakeholders to develop successful ML projects. In total, our approach outlines 28 tasks that are covered by the five perspectives. These tasks group associated concerns that should be analyzed by stakeholders. With this feature, stakeholders can more easily understand and describe how tasks are performed, what concerns are involved, the relationships between concerns, and the interactions with other stakeholders. For instance, typically in ML projects, data scientists are tasked with training, validating, and deploying ML models. These tasks involve implicit concerns that are not easily identified at first sight, such as inference time, learning time, model complexity and hyperparameters tuning. In addition, some specific tasks can benefit from involving more than one stakeholder in the analysis. For instance, to validate ML models it is necessary to generate model performance metrics, typically performed by data scientists, and analyze such metrics in collaboration with domain experts who deep understand the problem and data.

In the early phases of developing ML-enabled systems, several key tasks should be performed to lay a strong foundation for the project's success. These tasks typically involve all the stakeholders, and concern understanding the problem, setting goals, among other. Table 1 details the tasks from a system objectives perspective.

**Table 1** Description of the tasks to define the system objectives

| Task | Description |
| --- | --- |
| **Understand the problem** | understand the problem domain and the real-world context in which the ML model will be deployed, and define the ML problem and the specific task to be solved |
| **Set goals at different levels** | define the goals of the ML project at different levels in order to ensure that it meets the expectations of the stakeholders |
| **Establish success indicators** | define measures that provide early insights on the achievement of the objectives |
| **Manage expectations** | define what the ML model can and cannot do. Stakeholders may have unrealistic expectations about the ML capabilities, and providing clarity will prevent disappointment and frustration |

A positive user experience is crucial for the successful adoption, acceptance, and utilization of ML-enabled systems. It enhances user engagement, improves user satisfaction, and ultimately contributes to the overall success of the ML project. Table 2 details the tasks should be done to ensure that ML-enabled systems become a valuable and integral part of users' workflows.

**Table 2** Description of the tasks to ensure user experience

| Task | Description |
| --- | --- |
| **Establish the value of predictions** | determine that the ML model's outputs are relevant, accurate, and impactful and how they contribute to achieving the project's objectives |
| **Define the interaction of predictions with users** | define how users will interact with predictions (*e.g.*, frequency and forcefulness) in order to design user-friendly interfaces and workflows |
| **Visualize predictions** | present ML model outputs in a visually understandable format. Visual aids such as charts, and graphs can help users comprehend complex data and insights |
| **Collect learning feedback from users** | offer feedback mechanisms to users in order to provide updates on ML models |
| **Ensure the credibility of predictions** | ensure that users have a clear understanding of the ML model's capabilities and potential inaccuracies |

A robust and well-designed infrastructure is fundamental for the success of ML projects. It enables efficient development, deployment, and scaling of ML models. Table 3 details the tasks of the infrastructure perspective.

**Table 3** Description of the tasks to support the infrastructure of ML-enabled systems

| Task | Description |
|---|---|
| **Transport data to the model** | involves moving the relevant data from its source to the ML model for analysis, training, or prediction |
| **Make the ML model available** | refers to the process of deploying and exposing the trained ML model so that it can be accessed for making predictions |
| **Update the ML model** | refers to the process of making improvements or modifications to an existing ML model to enhance its performance |
| **Store ML artifacts** | involves the systematic storage and management of various artifacts generated throughout the ML development process |
| **Observe the ML model** | involves analyzing the performance, behavior, and outcomes of both the ML model and the software system |
| **Automate End-to-End ML workflow** | involves the design and implementation of a systematic and streamlined process that automates the ML workflow, from data preparation to model deployment and monitoring |
| **Integrate the ML model** | involves incorporating the trained ML model into the larger software system where it will be used for making predictions |
| **Evaluate the financial cost involved with infrastructure** | assess and analyze the expenses related to the computational resources, hardware, software, and services required to support the ML project |

A structured ML model development process fosters transparency, reproducibility, and accountability. It supports the creation of robust, reliable, and trustworthy ML solutions. Table 4 details the tasks of the model perspective.

**Table 4** Description of the tasks to support the creation of ML models

| Task | Description |
|---|---|
| **Select and configure the ML model** | shortlist a set of ML algorithms that are well-suited for the task at hand, and experiment with different combinations of hyperparameters to find the optimal configuration that yields the best performance |
| **Train the ML model** | create a ML model that captures the underlying patterns in the data and can make predictions on unseen examples |
| **Validate the ML model** | ensure that the trained ML model meets the desired criteria |
| **Deploy the ML model** | make the trained ML model available and operational in a production environment, allowing it to serve predictions to end-users or other systems |
| **Evaluate other quality characteristics** | assess various aspects of the ML model beyond its predictive accuracy. Other quality characteristics are equally important for the model's overall performance, reliability, and suitability for real-world applications |

The management of data in ML projects is essential for building accurate and reliable ML models. Table 5 details the tasks to be done, mainly by data scientists and domain experts, to maintain high-quality data throughout the lifecycle of ML projects.

**Table 5** Description of the tasks to support data quality in ML projects

| Task | Description |
|---|---|
| **Access data** | involves timely obtaining and retrieving the necessary data from various sources to be used for model development and evaluation |
| **Select and describe data** | involves carefully choosing the relevant data that will be used to train, validate, and test ML models, and describing the features of the data |
| **Evaluate high-quality data** | involves a comprehensive assessment of the data used for training and testing ML models in order to ensure that the data meets certain criteria and standards to produce accurate and reliable results |
| **Convert data in the representation of the ML model** | involves transforming the raw input data into a format that can be processed by the ML algorithm |
| **Split dataset** | involves dividing the available data into separate subsets for training, validation, and testing purposes |
| **Define a golden dataset** | involves creating a high-quality dataset that represents the problem domain and serves as the ground truth for training and evaluating ML models |

## 4.4 Perspectives

In SE, a perspective refers to a representation of a system or its components. It provides a focused way of analyzing a particular aspect of the system, allowing to capture different concerns and stakeholders' viewpoints. Perspectives have been effectively used in SE to model scenarios where team members work on a particular phenomena [5]. In *PerSpecML*, we modeled five perspective that are detailed as follows.

**System Objectives Perspective:** When evaluating ML solutions, there is a tendency to focus on improving ML metrics such as the F1-score and accuracy at the expense of ensuring business value and covering business requirements [4]. Success in ML-enabled systems is hard to define with a single metric, therefore it becomes necessary to define success at different levels. This perspective involves analyzing the context and problem that ML will address to ensure that ML is targeting at the right problem; defining measurable benefits ML is expected to bring to the organization and users; what system and model goals will be evaluated; the ML expected results in terms of functionality, and ML trade-off to deal with customer expectations. Table 6 details the concerns when thinking on objectives for ML-enabled systems.

**User Experience Perspective:** A good ML-enabled system includes building better experiences of using ML. The goal of this perspective is to present the predictions of the ML model to users in a way that achieves the system objectives and gets user feedback to improve the ML model. Therefore, we consider analyzing concerns such as defining what is the added value as perceived by users from the predictions to their work; how strongly the system forces the user to do what the ML model indicates; how often the ML model interacts with users; how the predictions will be presented so that users get value from them; how the users will provide new data for learning; and what is the user impact of a wrong ML model prediction. Table 7 details the concerns when thinking on user experience for ML-enabled systems.

Table 6 Description of each concern of the system objectives perspective

| Id | Concern | Addressing this concern involves specifying |
|---|---|---|
| O1 | Context | the specific circumstances, environment, or conditions in which the ML-enabled system will operate |
| O2 | Need | the requirement, desire, or gap that must be addressed to achieve a particular set of circumstances within a given context |
| O3 | ML functionality | the nature of the learning problem and the desired outcome that the ML model is designed to achieve (*e.g.*, classify customers) |
| O4 | Profit hypothesis | how the ML system's outcomes will translate into tangible gains for the organization |
| O5 | Organizational goals | measurable benefits ML is expected to bring to the organization. *E.g.*, increase the revenue in X%, increase the number of units sold in Y%, number of trees saved |
| O6 | System goals | what the system tries to achieve, with the support of an ML model, in terms of behavior or quality |
| O7 | User goals | what the users want to achieve by using ML. *E.g.*, for recommendation systems this could involve helping users find content they will enjoy |
| O8 | Model goals | metrics and acceptable measures the model should achieve (*e.g.*, for classification problems this could involve accuracy X%, precision Y%, recall Z%) |
| O9 | Leading indicators | measures correlating with future success, from the business' perspective. This includes the users' affective states when using the ML-enabled system (*e.g.*, customer sentiment and engagement) |
| O10 | ML trade-off | the balance of customer expectations (*e.g.*, inference time vs accuracy, false positive vs false negative) |

Table 7 Description of each concern of the user experience perspective

| Id | Concern | Addressing this concern involves specifying |
|---|---|---|
| U1 | Value | the added value as perceived by users from the predictions |
| U2 | Forcefulness | how strongly the system forces the user to do what the ML model indicates they should (*e.g.*, automatic or assisted actions) |
| U3 | Frequency | how often the system interacts with users (*e.g.*, whenever the user asks for it or whenever the system thinks the user will respond) |
| U4 | Visualization | user-friendly interfaces to showcase the ML model's outputs and facilitate its integration into the customer's existing systems (*e.g.*, specifying dashboard and visualization prototypes for validation) |
| U5 | Learning feedback | what interactions the users will have with the ML-enabled system in order to provide new data for learning, or human-in-the-loop systems where ML models require human interaction |
| U6 | Acceptance | how well and how the model arrives at its decisions |
| U7 | Accountability | who is responsible for unexpected model results |
| U8 | Cost | the user impact of a wrong ML model prediction |
| U9 | User education & Training | the need to provide user education and training on the limitations of the ML-enabled system and how to interpret its outputs |

**Infrastructure Perspective:** ML models produced by data scientists typically are turned into functional and connected software systems that demand special characteristics when in operation. The goal of this perspective is to cover the execution of the ML model, the monitoring of both data and model outputs, and its learning from new data. We consider analyzing concerns such as defining what streaming strategy will be used to connect data with the ML model; how the ML model will be served; the need for the ML model to continuously learn from new data to extend its knowledge; where the ML artifacts (*e.g.*, experiments, ML models, datasets) will be stored; the need for monitoring the ML model and data; the strategy to automate ML operations that allow to reproduce and maintain ML artifacts, and the integration the ML model will have with the rest of the system functionality. Table 8 details the concerns when thinking on the infrastructure for ML-enabled systems.

**Table 8** Description of each concern of the infrastructure perspective

| Id | Concern | Addressing this concern involves specifying |
|---|---|---|
| I1 | Data streaming | what data streaming strategy will be used (*e.g.*, real time data transportation or in batches) |
| I2 | Model serving | how the ML model will be executed and consumed (*e.g.*, client-side, back-end, cloud-based, web service end-point) |
| I3 | Incremental learning | the need for ML-enabled system abilities to continuously learn from new data, extending the existing model's knowledge |
| I4 | Storage | where the ML artifacts (*e.g.*, models, data, scripts) will be stored |
| I5 | Monitorability | the need to monitor the data and the outputs of the ML model to alert/detect when data drifts or changes |
| I6 | Telemetry | what ML-enabled system data needs to be collected. Telemetry involves collecting data such as clicks on particular buttons and could involve other usage data |
| I7 | Reproducibility | the need to repeatedly run an algorithm/ML process on certain datasets/experiments and obtain the same (or similar) results |
| I8 | Maintainability | the need to modify ML-enabled systems to improve performance or adapt to a changed environment |
| I9 | Integration | the integration that the model will have with the rest of the system functionality (*e.g.*, safety, security, privacy, fairness, legal) |
| I10 | Cost | the financial cost involved in executing the inferences and with the infrastructure that could affect architectural decisions. Great models can be unusable due to the cost to run and maintain them |

**Model Perspective:** Building a ML model implies not only cleaning and preparing data for analysis, and training an algorithm to predict some phenomenon. Several other aspects determine its quality. This perspective involves analyzing concerns such as defining the initial candidate of expected inputs and outcomes (of course, the set of meaningful inputs can be refined during pre-processing activities); the set of algorithms that could be used according to the problem to be addressed; the need to tune the hyperparameters of the algorithms; the metrics used to evaluate the ML model and measurable performance expectations that tend to degrade over time; the need for explaining and understanding reasons of the model outputs; the ability of the ML model to perform well as the size of the data and the complexity of the problem increase (scalability), to deal with discrimination and negative consequences for certain groups (bias & fairness), to protect sensitive data and prevents unauthorized access (security & privacy); the acceptable time to train and execute the ML model, and the complexity of the ML model in terms of size and generalization. In Table 9, we provide the description of the concerns that may be relevant to select, train, tune and validate a ML model.

Table 9 Description of each concern of the model perspective

| Id | Concern | Addressing this concern involves specifying |
|---|---|---|
| M1 | Algorithm & model selection | the set of algorithms that could be used/investigated, based on the ML problem and other concerns to be considered (*e.g.*, constraints regarding explainability or model performance, for instance, can limit the solution options) |
| M2 | Algorithm tuning | the need to choose a set of optimal hyperparameters for a learning algorithm. A hyperparameter is a parameter whose value is used to control the learning process |
| M3 | Input & Output | the expected inputs (features) and outcomes of the model. Of course, the set of meaningful inputs can be refined/improved during pre-processing activities, such as feature selection |
| M4 | Learning time | the acceptable time to train the model |
| M5 | Performance metrics | the metrics used to evaluate the model (*e.g.*, precision, recall, F1-score, mean square error) and measurable performance expectations |
| M6 | Baseline model | the optional simple model that acts as a reference. Its main function is to contextualize the results of trained models |
| M7 | Inference time | the acceptable time to execute the model and return the predictions |
| M8 | Model size | the size of the model in terms of storage and its complexity (*e.g.*, for decision trees there might be needs for pruning) |
| M9 | Performance degradation | the awareness of performance degradation. Over time many models' predictive performance decreases as a given model is tested on new datasets within rapidly evolving environments |
| M10 | Versioning | the versions of libraries, ensuring compatibility, and handling any conflicts or issues that may arise due to dependencies. This is important for maintaining reproducibility, portability, and ensuring that the ML model can be easily set up and executed on different systems |
| M11 | Interpretability & Explainability | the need to understand reasons for the model inferences. The model might need to be able to summarize the reasons for its decisions. Other related concerns such as transparency, may apply |
| M12 | Scalability | the need for the model to perform well as the size of the data and the complexity of the problem increases |
| M13 | Bias & Fairness | the need for the model to treat different groups of people or entities |
| M14 | Security & Privacy | the need for the model to protect sensitive data and prevents unauthorized access |

**Data Perspective:** Data is critical to ML. Poor data will result in inaccurate predictions. Hence, ML requires high-quality input data. Based on the Data Quality model defined in the standard ISO/IEC 25012 [25] and our own experience, we elaborate on the data perspective. In this perspective, we considered concerns such as defining from where the data will be obtained; the strategy to select data; the description of data; evaluating the inherent quality data attributes (*e.g.,* accuracy, completeness, consistency, real usage); what data operations and modeling must be applied; the expected data distributions and how data will be split into training, validating and test data; the time between when data is expected and when it is readily available for use, and the need for a golden dataset approved by a domain expert. Table 10 details the concerns when thinking on data for ML-enabled systems.

Table 10 Description of each concern of the data perspective

| Id | Concern | Addressing this concern involves specifying |
|---|---|---|
| D1 | Source | from where the data will be obtained |
| D2 | Timeliness | the time between when data is expected and when it is readily available for use |
| D3 | Data selection | the process of determining the appropriate data type and suitable samples to collect data |
| D4 | Data dictionary | the collection of the names, definitions, and attributes for data elements and models |
| D5 | Quantity | the expected amount of data according to the type of the problem and the complexity of the algorithm |
| D6 | Accuracy | the need to get correct data |
| D7 | Completeness | the need to get data containing sufficient observations of all situations where the model will operate |
| D8 | Credibility | the need to get true data that is believable and understandable by users |
| D9 | Real usage | the need to get real data representing the real problem |
| D10 | Bias | the need to get data fair samples and representative distributions |
| D11 | Consistency | the need to get consistent data in a specific context |
| D12 | Ethics | the need to get data to prevent adversely impacting society (*e.g.,* listing potential adverse impacts to be avoided) |
| D13 | Anonymization | the need to anonymize or pseudonymize to protect individual identities while still maintaining the utility of the data for ML purposes |
| D14 | Data operations & Modeling | what operations must be applied on the data (e.g., data cleaning and labeling) and what is necessary to convert data in the representation of the model. |
| D15 | Data distribution | the expected data distributions and how data will be split into training, validating and test data |
| D16 | Golden dataset | the need for a baseline dataset approved by a domain expert that reflects the problem. It is employed to monitor other data acquired afterwards |

## 4.5 Relationship between Concerns

Identifying relationships that show influence and implications between the concerns of an ML-enabled system is of paramount importance for successful project outcomes. These relationships extend across various dimensions, such as system design, risk management, and resource allocation. Understanding these factors allows for optimal decision-making, alignment with ML project goals, and efficient workflow planning.

In *PerSpecML*, we highlight these relationships to (i) help stakeholders identify conflicting objectives and requirements, and (ii) promote transparent communication between team members, ensuring the long-term viability and impact of ML projects. For instance, if users require to know the reasons of the ML model's decision-making then the explainability & interpretability concern arises. But this may depend on the chosen algorithm since some ML algorithms tend to be less explainable than others (*e.g.*, simpler ML algorithms such as decision trees, linear regression, and logistic regression are often considered more explainable than complex ML algorithms such as deep neural networks, random forests, and gradient boosting models). In addition, complex ML models may provide high accuracy, making it necessary to strike a balance between these concerns based on the specific needs and constraints of the ML project.

Identifying these relationships is also important within the infrastructure perspective. For instance, defining the source to access data influences the implementation or setup of a data streaming solution, which is required to transport the data to the ML model. Understanding these kind of relationships helps optimize the ML workflow and streamline the project execution. On the other hand, in the system objectives perspective, the ML functionality guides the selection of appropriate ML algorithms (*i.e.*, different tasks, such as classification or regression, require specific algorithms that are suitable for the task at hand). Furthermore, it affects how the ML model's performance is evaluated and measured (*i.e.*, different performance metrics, such as accuracy or recall are used based on the specific task). All *PerSpecML* relationships can be found in our online repository[1].

## 4.6 Perspective-Based ML Task and Concern Diagram

In order to provide a holistic view of the ML-enabled system that facilitates producing a description of what will be built and delivers it for approval and requirements management, we present a perspective-based ML task and concern diagram that integrates the key components discussed earlier: concerns, tasks, perspectives, and stakeholders. Table 11 shows the notation we used to represent these components in the diagram.

**Table 11** Legend of the perspective-based ML task and concern diagram

| Notation | Description |
|---|---|
| 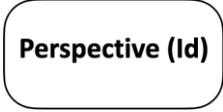 | The diagram contains five rounded rectangles that represent the perspectives. Each perspective is associated with a color to facilitate its identification, and is connected to their tasks |
| 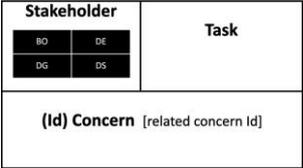 | The diagram contains rectangles attached to a perspective that connect a task (at the top right) to one or more concerns (at the bottom). Each task has at least one actor suggested (at the top left) related to the execution of the task and the analysis of the concerns |

---

[1]https://doi.org/10.5281/zenodo.7743479

The perspective-based ML task and concern diagram shown in Fig. 4 serves as a visual representation of the interplay between these components and their relationships within the context of ML projects. It offers a comprehensive overview of how different perspectives shape the tasks at hand, while considering the specific concerns associated with each task. Additionally, it highlights the involvement of various stakeholders who contribute their expertise and insights throughout the development process. By presenting this integrated diagram, we aim to provide a clear and structured approach for understanding the complex dynamics involved in building successful ML-enabled systems.

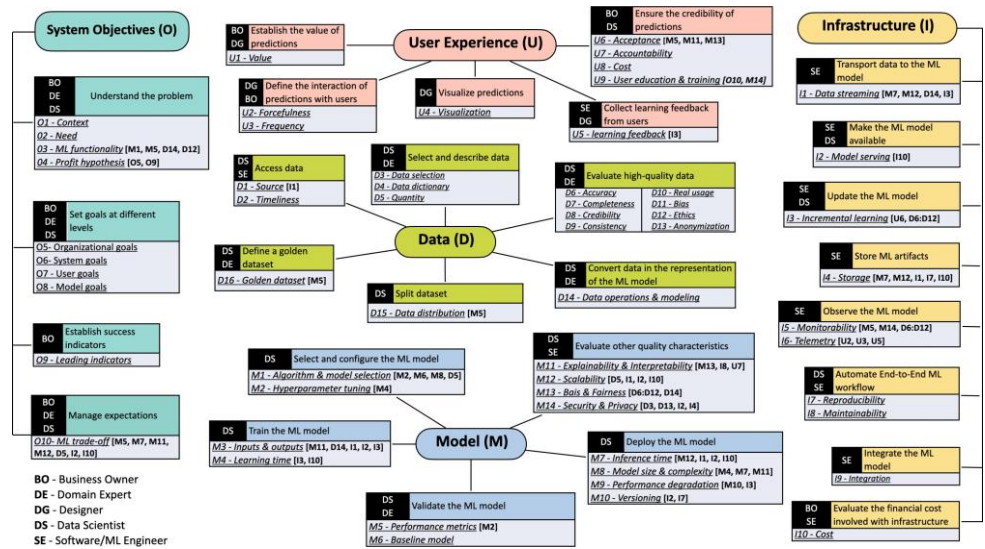

**Fig. 4** An illustration of the perspective-based ML task and concern diagram

## 4.7 Perspective-Based ML Specification Template

Documenting and organizing requirements is crucial for ensuring a clear understanding of the desired software system functionality, facilitating communication and collaboration, verifying and validating requirements, managing changes, and enabling knowledge transfer. It plays a vital role in successful software development and project outcomes. In order to fulfill these promises, we proposed a specification template based on the perspective-based ML task and concern diagram that provides a standardized format for documenting and organizing the applicable concerns of ML-enabled systems. Fig. 5 presents the perspective-based ML specification template for user experience and infrastructure perspectives. The complete template is available in our online repository[1]

Instead of starting from scratch each time, stakeholders can utilize this predefined template that already includes relevant sections, headings, and prompts, saving time and effort during the specification process. This reduces redundancy, and allows stakeholders to focus on the specific details and concerns of the ML-enabled system.

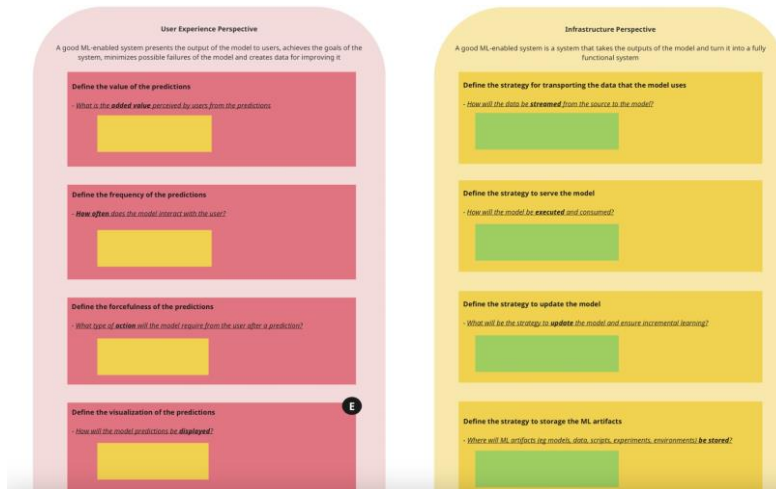

**Fig. 5** Perspective-based ML specification template for user experience and infrastructure perspectives

The perspective-based ML specification template consists of a set of predefined questions that guide the exploration and assessment of the concerns related to the tasks and perspectives. For example, if the concern is about the strategy to storage ML artifacts, the template includes a question that highlights ML artifacts such as models, data, experiment, and environments. If the concern is about the strategy to improve the performance of the ML algorithm, the template includes a question that highlights options such as hyper-parameter tuning. By analyzing these question-oriented concerns, we seek that stakeholders can ensure a comprehensive and systematic exploration of the concerns.

Inherently to the nature of ML projects, some types of concerns (*e.g.*, algorithm & model selection (M1) and data operations & modeling (D14)) are uncertain at the beginning of the project, mainly due to a common need of experimentation to get a better understanding on achievable requirements. Hence, they may be refined as the project progresses. The perspective-based ML specification template we proposed, highlights these concerns with the letter "E".

### 4.8 *PerSpecML'* Logical Flow

In order to provide clarity, structure, reproducibility, and consistency, this section shows the steps to be followed for executing *PerSpecML*. The purpose is to break down the overall process into manageable and sequential tasks, making it easier for stakeholders to understand and follow. Fig. 6 summarizes the logical flow to ensure that *PerSpecML* is executed in a systematic and organized manner, leading to more successful outcomes.

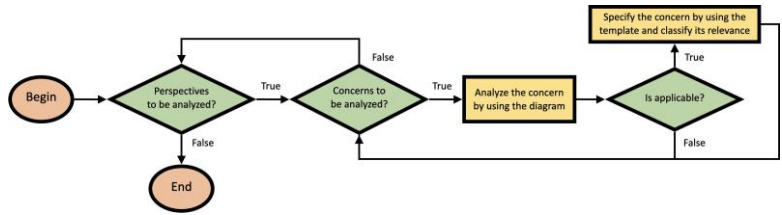

**Fig. 6** Logical flow for executing *PerSpecML*

We expect *PerSpecML* to be used by requirements engineers, in collaboration with other stakeholders, to support the specification of ML-enabled systems. The process begins by considering the perspectives. We established an intuitive order to analyze them: system objectives, user experience, infrastructure, model, and data. Given a perspective, a requirement engineer or a stakeholder performing that function can analyze each concern with the recommended stakeholders, also considering the relationships between concerns. If the concern is applicable, it should be specified in the perspective-based ML specification template and classify its relevance into desirable, important or essential.

## 5 Validation in Academia

As we mentioned before, *PerSpecML* is the result of a series of validations that were conducted in different contexts. The first validation was carried out within an academic environment where students had to use the candidate solution introduced in Section 3.3 to specify a toy problem. The simplified nature of the toy problem allowed for a clear understanding of how the candidate solution performed and how it could be improved. This led to valuable lessons and discoveries that were applied in the next validation with a more complex problem. In the following, we detail the validation in academia.

### 5.1 Context

The academic validation took place in the context of two courses on SE for data science with professionals, who are also students, from a Brazilian logistic company called Loggi[2], and computer science graduate students from the Pontifical Catholic University of Rio de Janeiro (PUC-Rio). Participants were asked to specify a feature for an ML-enabled system using the example of a bank loan problem, by analyzing the candidate solution's perspectives and concerns. The feature consisted of automatically classifying customers into good or bad payers and was described in user story format.

> *As a* Bank Manager *I want to* automatically classify customers *so that* I can decide upon granting a requested loan

From the user story, we can infer that the ML component needs to access, for learning purposes, data on customer characteristics, previously granted loans, and payment records. Regarding non-ML components and integration with other services, the participants could assume restrictions and requirements of the software system that the ML component would use. With this information, we asked the participants to analyze each concern of the candidate solution and provide a reasonable specification, if applicable. Thereafter they were asked to individually answer a follow-up questionnaire critically assessing the relevance and completeness of the candidate solution's perspectives and concerns. All the material provided to the participants is available in our online repository[1]. Fig. 7 illustrates the academic validation.

---

[2]https://www.loggi.com

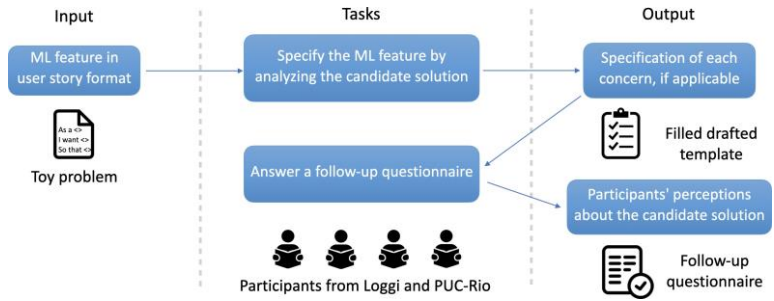

**Fig. 7** Process diagram for the academic validation

## 5.2 Goal and Method

We detail the goal of the validation in academia in Table 12. We followed the Goal-Question-Metric (GQM) goal definition template [5], which is a structured approach commonly used in SE and other disciplines, to help establish a clear connection between the overall goal, the specific questions that need to be answered, and the metrics used to measure progress.

**Table 12** Study goal definition of academic validation

| Analyze | the candidate solution's perspectives and concerns |
|---|---|
| for the purpose of | characterization |
| with respect to | perceived relevance and completeness, and ease of use, usefulness and intended use |
| from the viewpoint of | professionals and computer science graduate students |
| in the context of | two courses with 53 data science professionals from Loggi and 15 computer science students from PUC-Rio who were learning SE for data science |

Based on the goal, we established the following research questions for the validation in academia:

- **RQ1:** What is the relevance of each perspective of the candidate solution? We wanted to identify whether the perspectives of the candidate solution were perceived as meaningful and pertinent by the participants. This feedback helped confirm that the perspectives align with the needs and expectations of the intended users, and allowed us to identify areas that may need refinement.
- **RQ2:** Are the perspectives of the candidate solution and their concerns complete? This research question relates to the coverage of both the perspectives and concerns. This feedback helped to determine if critical components were missing or if there are gaps that need to be addressed.
- **RQ3:** To what extent does participants perceive the candidate solution as useful and beneficial? With this, we seek to understand the factors that influence the acceptance and adoption of the candidate solution. The question followed the technology acceptance model (TAM) [14] and aimed to capture participants' overall assessment and intention to use the candidate solution, incorporating elements of perceived usefulness, perceived ease of use, and intended use.
- **RQ4:** What are the limitations and opportunities for improvement of the candidate solution? This research question seeks feedback on the approach itself.

## 5.3 Selection of Subjects

The subjects were the attendants of two SE for data science courses. The in-company course at Loggi had 53 professionals with different background being trained in SE practices for building ML-enabled systems. The graduate course at PUC-Rio had 15 students (nine master and six Ph.D students). While students may have limited expertise compared to professionals in the field, they can provide fresh perspectives, helping us identify potential blind spots. In fact, using students as subjects remains a valid simplification of real-life settings needed in laboratory contexts [17]. In Table 13, we characterized the subjects by their educational background and average year of experience in ML projects.

Table 13  Subjects involved in the validation in academia

| Course | Total | Background | Experience (Average in years) |
|---|---|---|---|
| In-company | 33 | computer science | 1.2 |
|  | 20 | other discipline | 1.9 |
| University | 15 | computer science | 1.3 |

We can see that in the in-company course, not controlled by us, the professionals interested in data-driven projects are divided into those with a computer science background and those with background in other areas such as economics and mathematics. However, it is not surprising since the literature has already noted these findings for this role [28]. Overall, the participants were perceived as relatively inexperienced, as they possess only a few years of practical experience in developing ML-enabled systems. While the participants were selected by convenience (attendants of the courses), we believe that their profiles were suitable for our intended initial validation.

## 5.4 Data Collection and Analysis Procedures

To address the research questions related to the relevance, completeness, perceived usefulness, and potential improvements of the candidate solution in specifying ML-enabled systems, a questionnaire-based evaluation method was employed. This section outlines the data collection and analysis procedures used in the validation in academia.

**Questionnaire Design:** A follow-up questionnaire was designed to gather responses from participants regarding the research questions. The questionnaire included a combination of closed-ended questions related to *RQ1*, *RQ2* and *RQ3*, and one open-ended question related to *RQ4* to get both quantitative and qualitative data.

**Data Collection:** The questionnaire was delivered to the participants in online format for the in-company course and in-person session for the university course. Participants were provided with clear instructions on how to perform the specification task and how to complete the questionnaire and any specific considerations to keep in mind while responding.

**Quantitative Data Analysis:** For RQ1, RQ2, and RQ3, which involve assessing relevance, completeness, and perceived usefulness, quantitative data analysis techniques were employed. Closed-ended questions were used to capture participants' ratings on a two-point likert scale for *RQ1* and *RQ2*, and four-point likert scale for *RQ3*. Statistical analysis, such as mean and frequency distribution were computed to summarize the quantitative data.

**Qualitative Data Analysis:** For RQ4, which seeks to identify potential changes or additions to support practitioners, qualitative data analysis techniques were utilized. Open-ended questions allowed participants to provide detailed and descriptive responses. Qualitative analysis involved thematic coding, categorization, and identification of patterns or recurring themes across the responses.

**Interpretation and Findings:** The analysis of the collected data was interpreted according to the research questions. The findings were presented in a clear and concise manner, addressing each research question separately. In this case, charts were used to illustrate the results, providing a comprehensive overview of the validation in academia.

## 5.5 Results

### 5.5.1 RQ1. What is the relevance of each perspective of the candidate solution?

This question was designed as a single choice question. To assess the relevance of each perspective of the candidate solution, participants were asked to rate the importance high or low. The perspectives considered in this evaluation included ML objectives, user experience, infrastructure, model, and data. The results indicated that all perspectives were deemed relevant by the participants. Out of a total of 68 participants, 67 considered the data perspective highly relevant, indicating its significant importance in specifying ML-enabled systems. The ML objectives, model and infrastructure perspectives followed closely, at 66,65 and 63 respectively. The user experience perspective received a slightly lower number of 58, indicating its relatively high but somewhat lesser relevance. Fig. 8 presents the relevance of the candidate solution' perspectives based on their respective ratings.

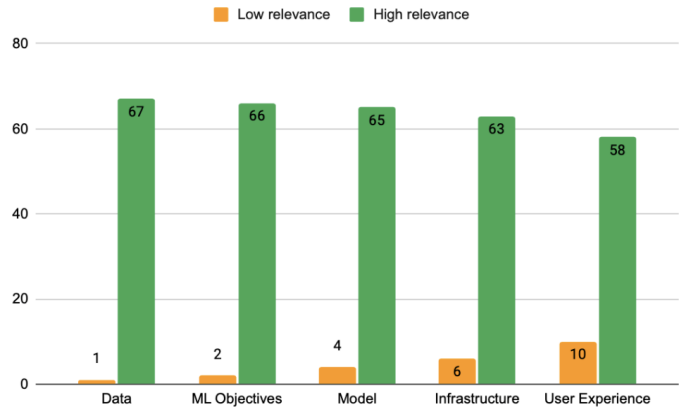

**Fig. 8** Frequencies of the relevance of each perspective of the candidate solution

Somehow we expect these results, since typically the main focus of practitioners in ML projects is data and models. In contrast, user experience concerns take a back seat to the development of ML-enabled systems. That is why with this work we seek to reinforce the importance of considering a user experience perspective.

### 5.5.2 RQ2. Are the perspectives of the candidate solution and their concerns complete?

This question was also designed as a single choice question with the option to explain the answer.To assess the completeness of perspectives and their associated concerns of the candidate solution, participants were provided with a list of predefined concerns corresponding to each perspective. They were then asked to indicate whether they believed the list was complete or if there were additional concerns that should be considered. The results revealed that participants generally considered the initial concerns and perspectives to be comprehensive but suggested some additional concerns. Only six out of 68 participants felt that something was missing. Across perspectives, the model perspective had the highest number of additional concerns identified by participants, highlighting the importance of monitoring ML models, optimizing parameters of ML algorithms, and breaking concepts about explainability. Below are the comments of the participants in that direction.

> "There should be a monitoring concern related to the model view. In the same way we have to train the model, we have to monitor the model outputs"

> "Parameter tuning in algorithms helps improve model performance. I would include this concern"

> "Explainability could be divided into two: explainability and interpretability, given that there are explainable models that are not necessarily interpretable"

### 5.5.3 RQ3. To what extent does participants perceive the candidate solution as useful and beneficial?

To gauge participants' perception of the acceptance of the candidate solution for specifying ML-enabled systems, participants were asked to rate the solution on various aspects. These aspects included ease of use, usefulness and intended use. Ratings were provided on a scale of 1 to 4 (four-point likert scale), with 1 indicating strongly disagree, 2 indicating partially disagree, 3 indicating partially agree, and 4 indicating strongly agree. The TAM questionnaire results are shown in Fig. 9.

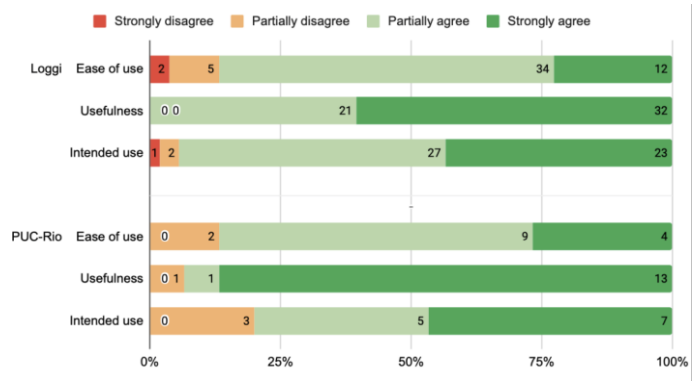

**Fig. 9** Frequencies of the TAM constructs for academic validation

The responses indicated a positive perception of the candidate solution. Participants from both courses rated the solution highly in terms of usefulness, with an average rating of 3.7, suggesting that the candidate solution can support the specification of ML-enabled systems. The ease of use of the candidate solution received an average rating of 3.1, indicating that the candidate solution did not provide enough guidance to be considered clear. The intended use of the candidate solution was rated at an average of 3.3, reflecting its feasibility and applicability. Overall, the candidate solution was perceived as highly useful, but showed potential for improvement in terms of ease of use. We understood that improving the candidate solution' guidance will imply an improvement in the perception of intended use.

### 5.5.4 RQ4. What are the limitations and opportunities for improvement of the candidate solution?

Here, participants had the option to respond in open text format. To identify potential improvements in supporting practitioners in specifying ML-enabled systems, participants were asked to provide suggestions regarding components, perspectives, or concerns that could be changed or added to enhance the candidate solution.The analysis of participants' responses revealed several valuable suggestions. As identified in the results of *RQ3*, some participants emphasized the need to further integrate the relationship between concerns. Others highlighted the importance of incorporating a road-map to apply the candidate solution. Additionally, one participant recommended providing more practical examples and case studies to enhance the solution's applicability. In the following, we present the comments of the participants in that direction.

> "It would be interesting to connect more concerns because I clearly see some relationships. For example, in the model perspective the explainability concern depends, to some extent, on the selection of the algorithm"

> "I would suggest explaining better how to use the approach because sometimes I did not know where to start and when to end"

> "Definitely a practical example would help to better understand the proposal"

These results provided insights into the relevance of the perspectives, the completeness of the concerns, the perceived usefulness, and potential improvements, guiding the refinement of the candidate solution. The validation in academia resulted in the following improvement opportunities.

1. In the infrastructure perspective, we decided to include '**monitorability**' as a new concern, since this may require implementing different services such as real-time logging, alerts, and data drift detection
2. In the model perspective, we broke the explainability concern into '**explainability and interpretability**', since these terms can have different interpretations
3. We added '**algorithm parameter tuning**' as a new concern of the model perspective, since data scientists typically need to analyze strategies to improve ML metrics
4. We defined a **set of steps** to be followed by stakeholders in order to apply the candidate solution

## 6 Static Validation In Industry

At this point, we made some improvements to the candidate solution, resulting in a version called *PerSpecML v1*. Building upon the foundation of the candidate solution, *PerSpecML v1* incorporates refinements and additions based on valuable feedback and insights from the students involved in the academic validation. In this section, we detail the second evaluation that was carried out in industry where practitioners had to use *PerSpecML v1* to retroactively specify two ready-made ML projects. We called this evaluation as static since it was performed without executing *PerSpecML v1* in a real or simulated environment.

### 6.1 Context

The static validation in industry involved practitioners of a R&D initiative called *ExACTa*[3] who developed two ML-enabled system projects from different domains for a large Brazilian oil company. The projects were developed following the Lean R&D approach [26] and are already deployed in production in several oil refineries. We refer to these projects as project A and B, since for reasons of confidentiality and undergoing patent requests they cannot be explicitly mentioned. Table 14 details these projects.

---
[3]https://exacta.inf.puc-rio.br

Table 14 Projects involved in the static validation

| Project | ML domain | Description |
| --- | --- | --- |
| A | Logistic regression | It alerts oil refineries about the likelihood of emitting strong odors that may result in claims from the community |
| B | Computer vision | It monitors images of the flame of oil refineries, helping refineries to decrease the disproportionate burning of gases that causes unnecessary energy consumption |

We retroactively specified Project A and B using *PerSpecML v1* with the support of the product owner of each project, analyzing the perspectives and their concerns, and filling a drafted specification template. This means that the specifications were added after the project had already finished. Thereafter, we discussed the resulting specifications in a focus group with the practitioners who developed these projects. Lastly, we provided to practitioners with a follow-up questionnaire to critically evaluate *PerSpecML v1*. All mentioned artifacts are available in our online repository[1]. Fig. 10 shows the process diagram for the static validation in industry.

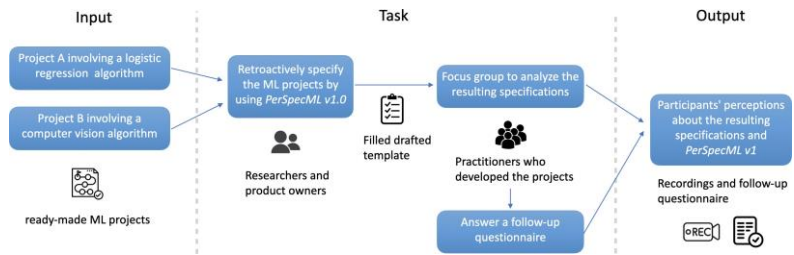

**Fig. 10** Process diagram for the static validation in industry

### 6.2 Goal and Method

We detail the goal of the static validation in Table 15. We followed the GQM template to describe what we evaluated in this first industrial validation. Here, we also describe the research questions.

In contrast with the academic validation, involving practitioners with more experience ensures the evaluation reflects real-world scenarios and challenges. Their expertise can provide valuable insights into the practical applicability of *PerSpecML v1* and its alignment with industry standards and best practices. Based on the goal, we established the following research questions for the static validation in industry.

**Table 15** Study goal definition of the static validation

| | |
|---|---|
| **Analyze** | *PerSpecML v1* (academically validated improved version) and its resulting specifications |
| **for the purpose of** | characterization |
| **with respect to** | perceived industrial relevance, ease of use, usefulness and intended use |
| **from the viewpoint of** | practitioners |
| **in the context of** | retroactively elaborated ML-enabled systems specifications using *PerSpecML v1* with six experienced software practitioners involved in the development of these systems |

- **RQ1:** What problems do participants face in practice when specifying ML-enabled systems? We wanted to identify the challenges and difficulties encountered by participants when specifying ML-enabled systems. By understanding these problems, we analyzed the adherence to our solution, and identified the suitability of *PerSpecML v1* to cover the needs of practitioners.
- **RQ2:** What perception do the participants have of the retroactive specifications of projects A and B derived from *PerSpecML v1*? By answering this research question, we gathered insights about the benefits or detriments of using *PerSpecML v1*.
- **RQ3:** What are the limitations and opportunities for improvement of *PerSpecML v1*? With the feedback received, we refined and enhanced *PerSpecML v1*
- **RQ4:** To what extent do the participants perceive *PerSpecML v1* as easy to use, useful and usable in the future? Through the TAM questionnaire, we explored the level of satisfaction and confidence participants had in *PerSpecML v1* as an approach for specifying ML-enabled systems.

## 6.3 Selection of Subjects

We invited six practitioners who have been actively working with the development of ML-enabled systems in the *ExACTa* initiative. Before starting the focus group and providing the questionnaire, we carefully selected the participants by asking them about the role and their experience in years working with ML projects. Table 16 shows an overview of the participant characterization.

**Table 16** Subjects involved in the static validation in industry

| Id | Role | Project | Experience (Years) |
|---|---|---|---|
| P1 | Data scientist | A | 6 |
| P2 | | B | 2 |
| P3 | | B | 2 |
| P4 | Developer | A | 2 |
| P5 | | B | 3 |
| P6 | Project lead | A | 2 |

It is possible to observe that in this study participants represent three different roles: data scientists who are interested in how the approach can help to build suitable and functional ML models, developers who are interested in how the approach can help to design the integration between components, and project leaders who are interested in how the approach can help the team achieve its goals. This allowed to gather feedback from people who have different needs and priorities. On the other hand, participants showed have more than two years of experience, helping us determine whether *PerSpecML v1* would work well in practice and what could be improved. Note that we selected three practitioners of each project involved in the evaluation.

## 6.4 Data Collection and Analysis Procedures

To address the research questions, a combination of focus group discussions and questionnaires were employed for data collection. In the following, we outline the data collection and analysis procedures used in the static validation in industry.

### 6.4.1 Focus Group

We conducted a focus group for promoting in-depth discussion on RQ1 and RQ2 [29]. Focus group is a qualitative research method that involves gathering a group of people together to discuss a particular topic, allowing for interaction between the participants, which can help to surface different viewpoints.

**Procedure:** The focus group was conducted in a structured and moderated format. The discussions were guided by the first author using open-ended questions related to RQ1 and RQ2, allowing participants to share their experiences, perspectives, and challenges faced when specifying ML-enabled systems.

**Data Collection:** We recorded the focus group with the consent of the participants to gather qualitative data. Transcripts of the focus group discussions were generated by the first author from the recordings, capturing participants' insights, ideas, and suggestions regarding RQ1 and RQ2.

**Data Analysis:** Thematic analysis was employed to identify common themes, patterns, and recurring topics in the focus group data [42]. The transcripts were coded, and emerging themes were categorized with the consensus of the authors. By last, the final set of categories were analyzed to address the research questions. The transcriptions and all codes are available in our online repository[1]. Examples of codes are highlighted when presenting the results.

### 6.4.2 Questionnaire

**Questionnaire design:** The questionnaire included structured questions and rating scales designed to capture quantitative and qualitative data related to RQ3 and RQ4, respectively. It addressed perceptions and feedback regarding the problems faced, usefulness of *PerSpecML v1*, ease of use, and identified limitations or opportunities for improvement.

**Data Collection:** The questionnaire responses were collected electronically through an online survey platform, taking care of anonymity and confidentiality. We provided the participants with clear definitions of the quality characteristics that we wanted to measure, ensuring that the participants understood what was asked of them.

**Data Analysis:** Quantitative data analysis techniques, such as descriptive statistics and inferential analysis, were used to analyze the questionnaire responses related to RQ4. These findings provided numerical insights and trends, allowing for a comprehensive understanding of participants' perceptions about the acceptance of *PerSpecML v1*. Qualitative data analysis techniques were also used to respond RQ3, involving coding and categorization.

## 6.5 Results

### 6.5.1 RQ1. What problems do participants face in practice when specifying ML-enabled systems?

We asked the participants about the problems they face when specifying ML-enabled systems. We coded and categorized the transcriptions of such discussions and then analyzed them to answer this research question. We found that participants frequently mentioned *lack of approaches to support the specification* given that ML incorporates additional challenges, which can make it difficult to specify ML-enabled systems. For instance, P6 stressed:

> "To the best of my knowledge there are no tools or approaches spread in industry helping practitioners to elicit, specify and validate requirements for ML systems"

In the same line, P4 and P5 complemented:

> "I'm curious to see a formal specification of an ML component. Based on my experience, these definitions are informal and emerge as the project progresses"

> "Sometimes I feel that the ML development team often transmits skepticism to customers, not because of the lack of knowledge of its members, but because of the lack of an established process to define what can be done in ML terms with what the customer makes available (*e.g.*, data, business information)"

On the other hand, we identified expressions about specification problems derived from the *need to involve domain experts*. For instance, P1 reported that understanding the specific domain plays a major role for accurate specifications:

> "Typically domain experts are busy, so they tend to be less involved in the early phases of ML projects. In the end, they often find unexpected results. Their involvement is important in areas such as feature engineering, data pre-processing and model evaluation"

P4 highlighted that customers often overestimate what ML can do. This leads to *unrealistic expectations of ML capabilities*, posing challenges in the specification process. The participant expressed:

> "Most of the time, customers expect that ML systems can solve all problems. They also don't imagine the number of components that are required to operate and maintain an ML model over time. Requirements engineering could help to address these challenges"

These findings reflect some of the problems faced by participants in practice when specifying ML-enabled systems, as identified through the focus group discussions with experienced practitioners. The insights gained from these discussions shed light on the key areas that require attention to overcome challenges such as *the lack of approaches to support the specification, the need to involve domain experts, and the customer unrealistic expectations of ML capabilities*

### 6.5.2 RQ2. What perception do the participants have of the retroactive specifications of projects A and B derived from *PerSpecML v1*?

After the participants analyzed the resulting specifications for Project A and B derived from *PerSpecML v1*, we asked them what they thought about it. Their feedback indicated positive perceptions of the specifications and their future impact on the development process. For instance, the participants highlighted that the specifications acted as a *guide during the development process*, helping to improve the overall development workflow. P1 manifested:

> "Looking at the diagram and its corresponding specifications allowed me to get an early overview of the requirements that can be refined as the project progresses. It is like a high-level guided development"

P1, P3 and P6 expressed that the retroactive specifications *enhanced clarity and understanding* of the ML-enabled systems for both projects:

> "I found that the specifications facilitated a better understanding of the systems' functionality, components, and data requirements, specially for Project A, in which I was involved"

> "I really liked the focus on diverse aspects such as data, model, and infrastructure. This landscape facilitates the understanding of the projects"

> "Identifying the tasks and concerns and their relationships allows identifying dependencies and influences as intended"

In addition, P3 mentioned that using *PerSpecML v1* allowed to *identify hidden concerns* that are not easily identified at first sight:

> "Typically, user experience concerns are put in the background. With *PerSpecML* was possible to early specify forcefulness, a concern analyzed late in the validation phase of Project B"

Finally, P5 noted that the retroactive specifications derived from *PerSpecML v1* helped in *documenting and communicating* the ML-enabled systems for both projects:

> "In my opinion, it is easy to convey the specifications to stakeholders, enabling better collaboration and alignment throughout the development process. For example, as a developer I can identify tasks where I need to collaborate with data scientists"

Overall, there was a clear consensus on the benefits of the retroactive specifications of Project A and B, derived from *PerSpecML v1*. According to the participants, the specifications *enhanced clarity and understanding, improved documentation and* *communication, acted as guide during the development process, and identified hidden concerns.*

### 6.5.3 RQ3. What are the limitations and opportunities for improvement of *PerSpecML v1*?

Participants' feedback revealed several limitations and opportunities for improvement. These insights, derived from the open-ended question of the questionnaire, can be related to the findings of RQ4, where we had participants who expressed partial agreement and disagreement about ease of use, usefulness, and intended use. For instance, P1 and P2 suggested that *providing additional guidance* could help users grasp *PerSpecML v1* more easily.

> "It is not clear to me how to get the specifications from analyzing the diagram. Even with the provided steps to apply the solution, it is not clear to me"

> "Providing tutorials or additional documentation could improve its application"

Participants also provided feedback on *improving the user interface* of *PerSpecML v1*, suggesting a more user-friendly design.

> "In my opinion, the specification template, which summarizes what the system should do, should be cleaner. I mean, the relationships between concerns are not needed as they exist in the diagram"

> "Better visualizations and intuitive navigation could further enhance the user experience and ease of use"

On the other hand, P6 commented on *improving the relationship between tasks and concerns*. More specifically, the participant suggested breaking down a task of the ML objective perspective, since the concerns were not related at all.

> "In the ML objective perspective there is something that does not make sense. The 'define objectives' task has independent concerns that could be part of separate tasks"

We identified limitations and opportunities for improvement of *PerSpecML v1* related to *providing additional guidance, improving the user interface, and improving the relationship between tasks and concerns*. Some of them may be related with the participants' perceptions explored in RQ4. We addressed these limitations and capitalized on the opportunities for improvement, allowing to refine *PerSpecML v1* to better meet the needs and challenges identified by practitioners.

### 6.5.4 RQ4. To what extent do the participants perceive *PerSpecML v1* as easy to use, useful and usable in the future?

The participants' responses to a TAM questionnaire indicated varying degrees of agreement or disagreement with statements about ease of use, usefulness, and intended use. While the majority of participants totally agreed with the statements, there were a few participants who expressed partial agreement or disagreement. More specifically, one participant encountered some difficulties in using *PerSpecML v1*, two participants had reservations about its usefulness, and one participant was not fully confident in using it in the future. The TAM questionnaire results are shown in Fig. 11.

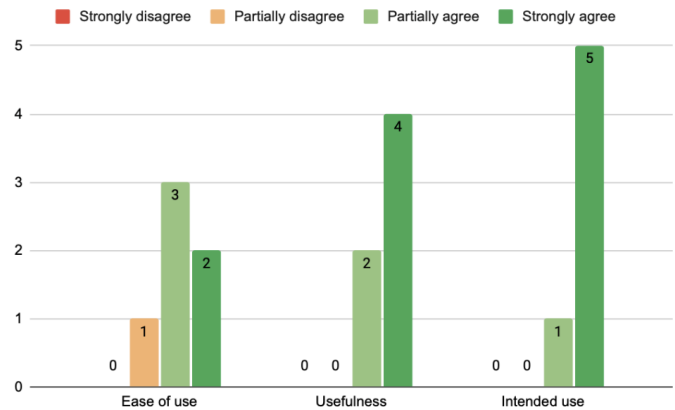

**Fig. 11** Frequencies of the TAM constructs for static validation industry

These varied perceptions explained to some extent the feedback received in RQ3 for identifying areas of improvement and addressing any concerns or challenges raised by participants. At the end of this validation, we decided to consider the feedback of the practitioners of the *ExACTa* initiative. In the following, we outline what was incorporated into *PerSpecML v1* from this static validation in industry.

> 5. We added the **domain expert role** to the *PerSpecML v1'* stakeholders, including it in tasks
> 6. The steps defined in the academic validation to apply *PerSpecML v1* turned into a **workflow diagram** to facilitate its application
> 7. We improved the *PerSpecML v1* documentation by creating a **Miro board**[a] that summarizes the perspectives, tasks and concerns to be analyzed. We also added a **practical use case** and **explanations** of each *PerSpecML* component
> 8. We improved the user interface of both diagram and specification template by adding **colors** that identify each perspective and their concerns
> 9. We simplified the specification template by **removing the representation of the relationships between concerns** (leaving them only in the perspective-based ML task and concern diagram, as they are used during the analysis)
> 10. We checked **terminology** and the **relationship between tasks and concerns** of each perspective to ensure its suitability
>
> ---
> [a]https://miro.com/miroverse/perspecml-machine-learning/

# 7 Dynamic Validation in Industry

Based on the valuable feedback and insights from the practitioners involved in the static validation, we made significant improvements to *PerSpecML v1*, resulting in a more robust and enhanced version called *PerSpecML v2*. In this section, we evaluated *PerSpecML v2* by performing (i) requirement workshop sessions and (ii) interviews with practitioners who work for a large Brazilian e-commerce company known as Americanas that offers technology, logistics, and consumer financing services. We called this validation as dynamic, since it was performed by executing *PerSpecML v2* for specifying two real ML projects from scratch.

## 7.1 Context

We conducted the dynamic validation on two distinct case studies at Americanas, where each case study involved a real ML-enabled system that was specified from scratch using *PerSpecML v2*. Each system was assigned a team made up of novice and experienced practitioners. The purpose of these ML-enabled systems is to enhance user experience, increasing engagement, and driving business goals of the Americanas company. Table 17 details the ML-enabled systems that were part of this evaluation.

Table 17 ML-enabled systems involved in the dynamic validation

| System | ML domain | Description |
|---|---|---|
| Product Classification | Natural Language Processing | It classifies titles of products registered by sellers in the marketplace of the Americanas company into categories. Based on the correct category, basic attributes for registering the product details are then provided to the seller |
| Market | Recommendation System | It suggests products to customers that are likely to be of interest or relevance to them. Based on historical data and similarity measures, the products are recommended |

Regarding the operation of these studies, we assisted practitioners in the application of *PerSpecML v2* in requirements workshop sessions by providing the necessary materials and information in advance. This included documentation on *PerSpecML v2* and example use cases. During the sessions, the practitioners analyzed and specified the ML-enabled systems by using *PerSpecML v2*. The specifications were made by adding post-its into the interactive Miro board we created from the static validation. Thereafter, we interviewed, in two additional sessions, the experienced practitioners who have knowledge of the domain problem and led the design and implementation of both ML-enabled systems to discuss the resulting specifications.Finally, we provided to all practitioners, a follow-up questionnaire to critically evaluate *PerSpecML v2* and the resulting specifications. All mentioned artifacts are available in our online repository[1]. Fig. 12 shows the process diagram for the dynamic validation in industry.

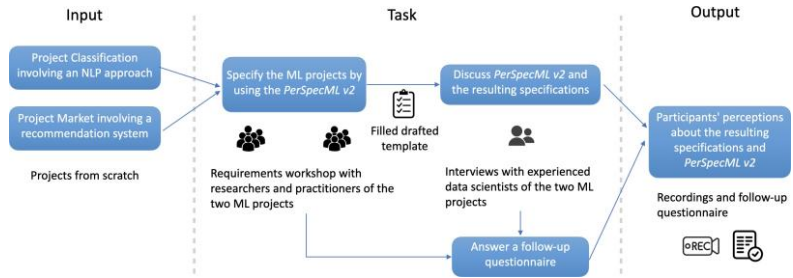

**Fig. 12** Process diagram for the dynamic validation in industry

## 7.2 Goal and Method

We detail the goal of the case studies of the dynamic validation in Table 18. We followed the GQM template to describe what we evaluated in this second industrial validation. Here, we also describe the research questions.

**Table 18** Study Goal Definition of the Dynamic Validation

| Analyze | *PerSpecML v2* (statically validated improved version) and its resulting specifications |
|---|---|
| **for the purpose of** | characterization |
| **with respect to** | the perceived quality of the specifications derived from *PerSpecML v2*, and ease of use, usefulness and intended use of *PerSpecML v2* |
| **from the viewpoint of** | practitioners |
| **in the context of** | two requirements workshop sessions involving 11 novice practitioners and three experienced practitioners who used *PerSpecML v2* to specify two ML projects from scratch, and (ii) two interviews with the three experienced practitioners who evaluated the resulting specifications derived from *PerSpecML v2* |

Based on the presented goal, aligned to the purpose of a dynamic industrial validation, we defined the following research question to better understand the practical suitability of using *PerSpecML v2*.

- **RQ1:** What perception do practitioners have while specifying ML-enabled systems by using *PerSpecML v2* ? For this research question, we conducted a comprehensive evaluation of practitioners' experiences while specifying ML-enabled systems using *PerSpecML v2*. During the requirements workshop sessions, we observed their interactions with *PerSpecML v2*, noted any challenges or difficulties they encountered, and gathered their feedback through discussions and direct feedback.
- **RQ2:** What perception do experienced practitioners have of the resulting specifications derived from *PerSpecML v2* ? To answer this question, we interviewed three experienced practitioners who reviewed and discussed the specifications derived from *PerSpecML v2*. We selected them since experienced practitioners can better assess the efficiency and effectiveness of *PerSpecML v2* than novice, for instance, by comparing it to existing methods they have used in the past. During the interview, the experienced practitioners provided their feedback and insights of the specifications. The goal was to gather valuable insights into how the experienced practitioners perceived the quality, completeness, and suitability of the specifications produced by using *PerSpecML v2.*

- **RQ3:** What are the limitations and opportunities for improvement of *PerSpecML v2* ? To explore this research question, we considered the feedback and discussions from both the novice and experienced practitioners. The novice practitioners' firsthand experience with using *PerSpecML v2* shed light on challenges, difficulties, and limitations they encountered while applying the approach. Additionally, the insights provided by the experienced practitioners allowed us to identify areas for improvement and potential enhancements. With the feedback received, we further refined *PerSpecML v2* and came up to its final version.
- **RQ4:** To what extent do the practitioners perceive *PerSpecML v2* as easy to use, useful and usable in the future? To address this research question, we provided to participants a follow-up questionnaire. We collected feedback from both the novice and experienced practitioners regarding their perception of *PerSpecML v2* as an approach for specifying ML-enabled systems. The novice practitioners, who used *PerSpecML v2* during the requirements workshop session, provided their insights on the ease of use, usefulness, and usability of the approach. Additionally, the experienced practitioners shared their opinions on the practicality and potential future utility of *PerSpecML v2*. By analyzing their feedback, we gained a comprehensive understanding of how *PerSpecML v2* was perceived by practitioners across different experience levels.

### 7.3 Selection of Subjects

The dynamic validation involved two main groups of participants from Americanas: novice practitioners who specified two ML-enabled systems from scratch using *PerSpecML v2*, and experienced practitioners who also specified the systems, and additionally evaluated the resulting specifications. The practitioners were characterized by having varied backgrounds, such as computer science, mathematics, physics, and others. The diversity in their educational background and experience helped validate the maturity of *PerSpecML v2*. Their feedback shed light on its suitability for real-world implementation and if it meets the expectations and requirements of industry professionals. In Table 19, we characterized the subjects by their role in the development of the ML-enabled systems involved in this study, educational background, and years of experience involved in ML projects.

The subjects involved in specifying the ML-enabled systems from scratch were divided into two teams. In the fist one that we call team A, we had six novice practitioners and one experienced practitioner responsible for *Product classification* system. In the second team that we call B, we had five novice practitioners and two experienced practitioners responsible for *Market* system. We highlighted the experienced practitioners who led each team with grey color in order to differentiate them from novice. Note that experienced practitioners are data scientists with a different educational background than computer science or engineering (except for P14), as expected for these positions [3, 28].

**Table 19** Subjects involved in the dynamic validation in industry

| Team | Id | Role | Background | Experience (Years) |
|---|---|---|---|---|
| Team A | P1 | Developer | Computer science | 1 |
| | P2 | | Design | 1 |
| | P3 | | Computer science | 1.5 |
| | P4 | | Computer engineering | 1 |
| | P5 | Scrum master | Physics | 1.5 |
| | P6 | Data scientist | Computer science | 1 |
| | P7 | Data scientist | Linguistic | 8 |
| Team B | P8 | Developer | Electronic engineering | 1 |
| | P9 | | Computer engineering | 1 |
| | P10 | | Computer science | 1 |
| | P11 | | Mathematics | 1 |
| | P12 | Scrum master | Computer science | 2 |
| | P13 | Data scientist | Electrical engineering | 4 |
| | P14 | Data scientist | Computer science | 6 |

## 7.4 Data Collection and Analysis Procedures

To address the research questions outlined in this dynamic validation, we employed three main data collection procedures: requirements workshop sessions, interviews, and a follow-up questionnaire.

### 7.4.1 Requirements Workshop Sessions

**Workshop Design:** We designed the requirements workshop sessions with a clear agenda and objectives, and outlined the tasks that the participants performed during the workshop, such as using *PerSpecML v2* to specify the two ML-enabled systems from scratch. This allowed to provide the input to respond to RQ1.

**Data Collection:** During the sessions, we collected data in the form of written specifications produced by the practitioners. These specifications included concerns on the five perspectives such as objectives, user experience, infrastructure, model, and data.

## 7.5 Interviews

**Interview Design:** We developed a semi-structured interview protocol for RQ1. The protocol included a set of open-ended questions that focus on the experienced practitioners' perception of the resulting specifications derived from *PerSpecML v2*. Questions explored aspects such as the quality, completeness, clarity, and effectiveness of the specifications. This shed light on answering RQ1.

**Data Collection:** We conducted interviews with the experienced practitioners. During the interviews, we used the protocol to guide the discussions, while allowing practitioners to share their thoughts and insights freely. We recorded the interviews in video format, with their consent, in order to ensure accurate capture of responses and allows for later review and analysis.

**Data Analysis:** We transcribed the video recordings of the interviews into text format in order to analyze the participants' responses, and then we applied coding techniques to categorize them into themes. In addition, we triangulated by comparing and cross-referencing the results from the different interviewees.

**Reporting:** We summarized the findings and insights from the interviews in a structured manner by including direct quotes and paraphrased statements from the practitioners to support the analysis and interpretations.

## 7.6 Questionnaire

**Questionnaire design:** The questionnaire included structured questions and rating scales designed to capture quantitative and qualitative data related to RQ2 and RQ3, respectively. It addressed perceptions and feedback regarding the usefulness and ease of use of *PerSpecML v2*, and identified limitations or opportunities for improvement.

**Data Collection:** The questionnaire responses were collected electronically through an online survey platform, taking care of anonymity and confidentiality.

**Data Analysis:** Quantitative data analysis techniques, such as descriptive statistics and inferential analysis, were used to analyze the questionnaire responses related to RQ2. These findings provided numerical insights and trends, allowing for a comprehensive understanding of participants' perceptions about the acceptance of *PerSpecML v2*. Qualitative data analysis techniques were also used to respond RQ3, involving coding and categorization.

## 7.7 Results

### 7.7.1 RQ1. What perception do practitioners have while specifying ML-enabled systems by using *PerSpecML v2*?

During the workshop specification sessions, we observed the interactions of practitioners with *PerSpecML v2* to identify benefits or difficulties they encountered. The comments and discussions indicated that practitioners had a generally positive perception of *PerSpecML v2* as a supportive tool for guiding them through the specification process. For instance, novice practitioners P3 and P5 appreciated *the visual and intuitive interface of PerSpecML v2*:

> "At first sight, I was able to identify each perspective, its tasks, and their concerns. This helps me to better understand the requirements and dependencies of the *Product Classification* system"

> "I find the specification template and language constructs within *PerSpecML* beneficial in structuring the specifications effectively"

As the workshops progressed, practitioners recognized the *PerSpecML v2*'s role in *early identification and resolution of potential concerns* in ML projects, and its *ability to facilitate collaboration and communication* among different teams involved in ML projects. P11, P13, P1 and P3 expressed:

> "Many times in our projects some of these concerns are only addressed when it is clearly too late. I see the diagram as a roadmap that allows me to identify components that would not be identified without its use"

> "There are several tasks that at the beginning of the project do not concern our team, but that deserve to be analyzed for their relationships with others"

> "*PerSpecML* summarizes the work of several ML teams in one diagram"

> "Linking the model update task in the infrastructure perspective with the need to get user feedback in the user experience perspective makes sense. This encourages communication between teams involved in ML projects"

While some initial learning curve was observed, practitioners quickly grasped *PerSpecML v2*'s functionalities and became comfortable using the approach. Their perception of usability and effectiveness improved as they gained more hands-on experience during the workshop sessions. RQ3 gave us more insights in this line.

### 7.7.2 RQ2. What perception do experienced practitioners have of the resulting specifications derived from *PerSpecML v2*?

The experienced practitioners expressed positive feedback regarding the resulting specifications derived from *PerSpecML v2* for the two ML projects. For instance, P13 and P14 appreciated the *clear and well-structured nature of the specifications*, and the *utility for specific users*:

> "The specifications demonstrated a good understanding of the ML projects' requirements, guiding the novice practitioners through the specification process"

> "The diagram can be extremely helpful for novice data scientists or engineers to get an overview of the ML workflow"

However, P7 pointed out minor areas where specifications could be further refined to better align with specific project needs:

> "I am not sure if at the end the specifications are already sufficiently clear, but I can state what has been raised is reasonable and useful. Coming up with a clear specification requires refinements and increments"

Indeed, the requirements workshop was supposed to be the first effort towards comprehensive specifications that should be further improved after the workshop. On the other hand, P7 and P14 (experienced practitioners from separate workshops) both compared *PerSpecML v2* with the approach they used so far in their projects.

> "*PerSpecML* provides a more comprehensive overview and is far better than the ML canvas to support specifying ML-enabled systems"

> "Currently, we use *ML canvas* to describe ML systems, but *PerSpecML* covers more elements, and helps analyze their relationships"

Overall, the experienced practitioners were impressed with the novice practitioners' efforts and saw *PerSpecML v2* as a valuable tool for fostering collaboration and understanding between different skill levels within the team.

### 7.7.3 RQ3. What are the limitations and opportunities for improvement of *PerSpecML v2*?

The open-ended responses in the follow-up questionnaire provided valuable insights into the limitations and opportunities for improvement of *PerSpecML v2*. For instance, P7 suggested adding a concern related to the *financial cost* associated with the infrastructure that is required to operate an ML-enabled system, while P3 recommended paying attention to the *versioning of libraries*.

> "Based on my experience, ML systems can be expensive to maintain. Even large companies should carefully consider the costs of maintaining ML systems before implementing them. I would include this concern for sure"

> "It is important to consider the versioning of the libraries that are typically used in the development of ML-enabled systems. On several occasions I have seen my teammates in trouble, for example, when the Python version is not compatible with the TensorFlow version. If there is a proper version management this could be avoided"

Moreover, P13 suggested complementing the model perspective with the phenomenon that occurs when the performance of ML models decreases over time, and that both data scientists and customers typically pass up.

> "Requirements specifications captures what the system is supposed to do, right? ML models tend to degrade over time due to several factors such as environmental and data changes. This behavior is typically not considered, therefore, it should be specified"

On the other hand, P12 added another interesting opportunity for improvement: classifying the concerns by importance to better cope with the number of concerns to be analyzed.

> "When analyzing the diagram I see that the number of concerns is considerable. That's not a bad, in fact, it shows everything to think when designing ML systems. For this reason, I think it would be interesting to classify each concern by its importance. This would somehow prioritize the specification process"

Finally, P14 mentioned the importance of automating *PerSpecML v2*:

> "It would be good to automate the approach by decreasing human involvement in the execution of *PerSpecML* that are prone to errors. It is a matter of practicality. In short, you can automate the *PerSpecML*' logical flow"

Overall, the feedback indicated that *PerSpecML v2* had potential for enhancement, and practitioners were eager to see future updates and features that could further elevate the tool's usability and effectiveness.

### 7.7.4 RQ4. To what extent do the practitioners perceive *PerSpecML v2* as easy to use, useful and usable in the future?

Based on the TAM questionnaire that included four-point Likert scale ratings, we found that practitioners indicated a high level of acceptance and positive perception of *PerSpecML v2*. The summary of the responses is shown in Fig. 13.

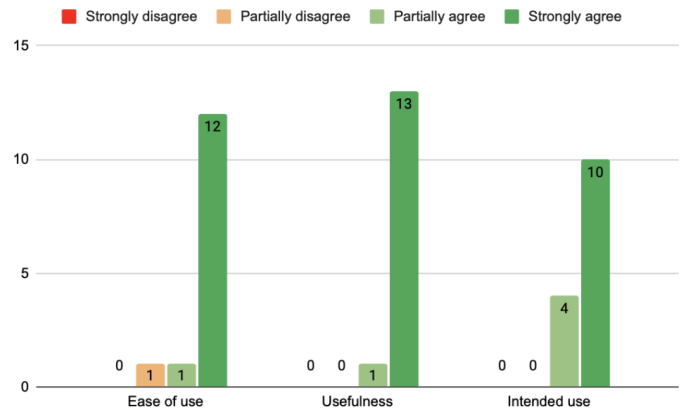

**Fig. 13** Frequencies of the TAM constructs for dynamic validation in industry

The majority of participants rated *PerSpecML v2* as easy to use, with a significant portion (12 out of 14) giving it a rating of 4 (strongly agree). The documentation, intuitive interface and clear instructions provided by *PerSpecML v2* –improvements that came up in static validation–contributed to its perceived ease of use, making it accessible and user-friendly for both novice and experienced practitioners. However, one participant expressed partial disagreement with the statement of ease of use. This response came from P14, an experienced data scientist who mentioned suggestions for improvements on this topic in the previous question.

Additionally, the practitioners found *PerSpecML v2* to be highly useful in the specification process. Excluding one who expressed partial agreement, all the participants gave it a rating of 4 for usefulness (strongly agree). Indeed, the discussions and the outputs of the workshop sessions showed that *PerSpecML v2* was especially valuable in guiding practitioners through the specification process and enhancing the overall clarity of the specifications.

Furthermore, the practitioners showed positive attitudes towards *PerSpecML v2* 's intended use. The majority of respondents (10 out of 14) expressed that they would be willing to use *PerSpecML v2* in future ML projects, indicating the approach's potential to become an essential part of their workflow for specifying ML-enabled systems.

Overall, the questionnaire results demonstrated a strong acceptance and positive perception of *PerSpecML v2* 's ease of use, usefulness, and future usability among the practitioners. When comparing these results with the static validation, we saw that the perception of ease of use improved considerably, indicating that the improvements from that evaluation had an effect.

At the end of this validation, we decided to consider the feedback of the practitioners of the Americanas company. In the following, we outline what was incorporated into *PerSpecML v2* from this dynamic validation in industry, which led to the final version of *PerSpecML*.

> 11. We added '**financial cost**' as a new concern of the infrastructure perspective, since ML typically demand implementing several services that impact project budget
> 12. We added '**versioning**' as a new concern of the model perspective, since this is essential for reproducibility, compatibility, and long-term maintainability of ML models
> 13. We added '**performance degradation**' as a new concern of the model perspective, since it can lead to inaccurate predictions, which can cause problems for businesses and organizations
> 14. Based on a meta-review of the validations, we included '**education & training**' in the user experience perspective, and '**anonymization**' in the data perspective. The first new concern will help that users have a clear understanding of the ML model's capabilities and potential inaccuracies ensure the system's credibility and user satisfaction, and the second one will help to protect sensitive data when required while still maintaining the utility of the data for ML purposes
> 15. We refined the *PerSpecML v2 '* **logical flow** to explicitly include the relevance of the concerns into desirable, important or essential. This could help to prioritize the requirements of ML-enabled systems

## 8 Threats to Validity

Assessing the validity of study results is particularly important for ensuring the accuracy, reliability, and generalization of findings. In this study, we empirically evaluated *PerSpecML* by analyzing human factors, such as practitioners' perceptions and experiences. In the following, we critically examine potential limitations and challenges that could impact the trustworthiness and applicability of our research outcomes. To this end, we followed the categories suggested by Wohlin *et al* [50].

**Construct validity:** For our quantitative and qualitative analyses, we conducted a mix of data collection methods, such as the TAM questionnaire, focus groups, and interviews. These choices were based on the well-established theoretical foundation of such methods. For instance, the TAM model has been widely used in technology acceptance research [44], and its questions were carefully designed to measure specific constructs related to the users' attitudes and intentions towards adopting our approach.

**Internal validity:** In the static validation, the practitioners' familiarity with the ML projects that were retroactively specified may have influenced their perception and performance during the validation process, leading to potential bias in the results. To mitigate this threat, we decided to retroactively specify the ML projects with the support of the product owner of each project, but without involving the practitioners. In this case, we wanted to take advantage of this situation since by knowing the ML projects, the practitioners could more easily evaluate the resulting specifications, *e.g.*, whether important aspects was missing.

**External validity:** We are aware that the generalization of the findings from the academic and static validation to real-world industrial scenarios may be limited. For instance, the toy scenario used in the academic setting and the specifications built retroactively in the static validation may not fully capture the complexity and challenges faced in actual industrial projects. Our intention with these artifacts was to use them to iteratively improve *PerSpecML* until it was mature and could be evaluated in a more realistic setting. Regarding the subject representativeness, we believe that the validation conducted in academia with students, and in industry with novice and experienced practitioners, constitutes a diverse setting that allowed for the examination of *PerSpecML* across different scenarios, thereby strengthening the generalization of the findings.

**Conclusion validity:** During the data collection and analysis procedures of the three evaluations, we used a single researcher for open coding. To mitigate this threat, we peer-reviewed the list of codes attached to the transcriptions, and validated our findings with the participants of the academic, static and dynamic validation. Therefore, as suggested by [29], we presented our conclusions to the involved participants to validate their agreement. Moreover, we triangulated both qualitative and quantitative data helped provide a more robust understanding of *PerSpecML*'s usability and effectiveness, supporting well-informed conclusions.

## 9 Discussion

In this section, we reflect on the outcomes of the validations and how they contribute to the understanding and improvement of *PerSpecML*, our perspective-based approach for specifying ML-enabled systems. We explore the broader implications of the findings, other areas of study, and how our approach can positively impact the development of ML-enabled systems.

In terms of **rigor**, *PerSpecML* is the result of a series of validations that were conducted in different contexts, each contributing valuable insights and refining our approach to meet the diverse needs of practitioners involved in ML projects. Through careful evaluations encompassing academia and industry, *PerSpecML* has undergone iterative enhancements, ensuring its effectiveness and adaptability in guiding the specification of ML-enabled systems across various scenarios and project complexities. The combination of student validation, real-world discussions with experienced data scientists, and collaborative evaluations with both novice and experienced practitioners has culminated in a robust and user-friendly approach that empowers teams to collaboratively and comprehensively define ML-enabled systems from inception to completion.

In terms of **scope and coverage**, *PerSpecML* was designed with the underlying assumption that the problem to be solved can benefit from ML, which is not always the case. Guidance to assess this assumption is out of our scope. While the focus of *PerSpecML* are requirements engineers, the specialists who provide a clear understanding of what needs to be built, other stakeholders such as project leaders can preside the application of *PerSpecML*. In addition, we are aware that not every ML-enabled system needs to address all the concerns we proposed and not every ML-enabled system needs to implement them to the same degree. Beyond qualities of ML components, of course, we also care about qualities of the system as a whole, including response time, safety, security, and usability. That is, traditional RE for the entire system and its non-ML components is just as important. Note that when considering the overall system, general quality characteristics of software products such as the ones mentioned in the ISO/IEC 25010 standard [24], should also be analyzed.

In terms of **expected benefits**, the main purpose of *PerSpecML* is to support the specification of ML-enabled systems by analyzing the ML perspective-based diagram and filling out the ML specification template. Nevertheless, we believe *PerSpecML* may eventually be useful in various situations. First, to validate an already specified ML-enabled system. In this case, the concerns would be a reference since they came from diverse source of knowledge (literature review, practical experiences and an external industrial experience on building ML-enabled systems [22]). Second, *PerSpecML* may help design ML-enabled systems, since it includes (i) different components, including functional and non-functional properties, (ii) how they interact with each other, (iii) how they are deployed, and (iv) how they contribute with business requirements. Third, *PerSpecML* is applicable to the most common ML approaches from typical ML domains, such as classification or regression problems, to more complex domains, such as computer vision and natural language processing. In fact, in the validations we conducted, we used different type of ML domains.

## 10 Concluding Remarks

In this paper we presented *PerSpecML*, a perspective-based approach for specifying ML-enabled systems, designed to identify which attributes, including ML and non-ML, are important to contribute to the overall system's quality. The approach empowers requirements engineers to analyze, with the support of business owners, domain experts, designers, software and ML engineers, and data scientists, 59 concerns related to typical tasks that such practitioners face, grouping them into five perspectives: system objectives, user experience, infrastructure, model, and data.

We introduced two main artifacts of *PerSpecML*: (i) the perspective-based ML tasks and concern diagram that provides a holistic view of ML-enabled systems, and (ii) its corresponding specification template that provides a standardized format for documenting and organizing the applicable concerns. Together, these artifacts serve to guide practitioners in collaboratively and comprehensively designing ML-enabled systems, enhancing their clarity, exploring trade-offs between conflicting requirements, uncovering hidden or overlooked requirements, and improving decision-making.

The creation of *PerSpecML* involved a series of validations conducted in diverse contexts, encompassing both academic and real-world scenarios as suggested in [20] for scaling proposals up to practice. The evaluation process began with a validation in academia, where students from two courses of SE for data science participated in specifying an ML-enabled system for a toy problem. This initial validation mainly showcased the promise of the approach and its potential for improvement in terms of ease of use. The static validation in an industry setting involved discussions with practitioners of a R&D initiative, analyzing specifications retroactively for two ready-made ML projects. This validation highlighted *PerSpecML*'s role as a roadmap for identifying key components that could be missed without using the approach, but also identified opportunities for improvements related to usability. Lastly, the dynamic validation engaged both novice and experienced practitioners of a Brazilian large e-commerce company, who specified two real ML-enabled systems from scratch using *PerSpecML*. The feedback from previous validations allowed the practitioners to focus on improvements related to the completeness of the concerns and how to use the approach. As a result of the diverse evaluations and continuous improvements, *PerSpecML* stands as a promising approach, poised to positively impact the specification of ML-enabled systems.

While the validations of *PerSpecML* have yielded promising results and provided valuable insights, there remain several avenues for future work and enhancements to further enrich the approach and its applications in the field. For instance, investigating ways to automatically generate detailed documentation from the specifications provided in *PerSpecML* artifacts could significantly streamline project management and maintainability. This would further bridge the gap between specification and implementation phases. In addition, conducting other studies and soliciting continuous feedback from practitioners who actively use *PerSpecML* in real projects would offer valuable insights into its long-term benefits. By last, given the potentially conflicting nature of the concerns highlighted in *PerSpecML*, delving into the study of trade-offs becomes even more promising, as it offers a pathway to address the complex particularities of ML-enabled systems.

## Acknowledgment

We would like to thank the employees of Loggi, of the ExACTa Initiative at PUC-Rio and of Americanas S.A. Thanks also for the financial support of the Brazilian CAPES and CNPq agencies (grant 312827/2020-2).